%% file: arXiv.tex
\documentclass[11pt]{article}
\usepackage[a4paper,margin=1.2in]{geometry}
\usepackage[font=small]{caption}
\usepackage{authblk}
\usepackage{setspace}
\usepackage{amsthm}
\usepackage{amsmath}
\usepackage{amsfonts}
\usepackage{amssymb}
\usepackage{bbm}
\usepackage{bm}
\usepackage{booktabs}
\usepackage{color}
\usepackage{enumerate}
\usepackage{framed}
\usepackage{float}
\usepackage[bottom,symbol]{footmisc} 
\usepackage{mathtools}
\mathtoolsset{showonlyrefs}
\usepackage{multirow} 
\usepackage{url} 
\usepackage{xr}    
\usepackage{appendix}
\usepackage{algorithm}
\usepackage{algpseudocode}

\input{preamble}

\renewcommand{\bx}{\mathbf{x}}
\renewcommand{\by}{\mathbf{y}} 
\renewcommand{\bX}{\mathbf{X}}
\renewcommand{\bK}{\mathbf{K}} 
 
\newcommand{\dd}{\mathrm{d}}

\RequirePackage{amsthm,amsmath,amsfonts,amssymb}
\RequirePackage[authoryear]{natbib}
\RequirePackage[colorlinks,citecolor=blue,urlcolor=blue]{hyperref}
\RequirePackage{graphicx}

\theoremstyle{plain}

\theoremstyle{definition}

\title{Bayesian Nonlinear PDE Inference via Gaussian Process Collocation with  Application to the Richards Equation}
\author{Yumo Yang$^1$, Anass Ben Bouazza$^2$, Xuejun Dong$^3$ and Quan Zhou$^{1,}$\footnote{Corresponding author: quan@stat.tamu.edu} \medskip \\   $^1$Department of Statistics, Texas A\&M University  \\  
$^2$Department of Statistics, University of Oxford\\
$^3$Department of Soil \& Crop Sciences, Texas A\&M University}  
\date{}

\begin{document}
\maketitle
\vspace{-2em} 
\begin{abstract}
The estimation of unknown parameters in nonlinear partial differential equations (PDEs) offers valuable insights across a wide range of scientific domains. In this work, we focus on estimating plant root parameters in the Richards equation, which is essential for understanding the soil-plant system in agricultural studies. 
Since conventional methods are computationally intensive and often yield unstable estimates,  we develop a  new Gaussian process collocation method for efficient Bayesian inference. 
Unlike existing Gaussian process-based approaches, our method constructs an approximate posterior distribution using samples drawn from a Gaussian process model fitted to the observed data, which does not require any structural assumption about the underlying PDE.  
Further, we propose to use an importance sampling procedure to correct for the discrepancy between the approximate and  true posterior distributions. As an alternative, we also devise a prior-guided Bayesian optimization algorithm leveraging the approximate posterior. 
Simulation studies demonstrate that our method yields robust estimates under various settings. 
Finally, we apply our method on a real agricultural data set and estimate the plant root parameters with uncertainty quantification.  
\end{abstract}

\section{Introduction} \label{sec:intro} 
Mathematical modeling of soil water dynamics and quantification of plant root water uptake play a central role in agricultural science,   serving as the foundation for understanding
the environmental impacts of soil-plant system and optimizing crop productivity~\citep{Wang2004}. 
To capture the temporal and spatial movement of water in unsaturated soil, Richards equation is widely used~\citep{BARRY199329,Dong2010}, 
which is a highly nonlinear partial differential equation (PDE)  relating soil water content and its derivatives via soil parameters~\citep{FEDDES1976}. 
A sink term can be added to the equation to explicitly model the water taken by plant root, but it often involves plant root parameters that are difficult to estimate~\citep{Dong2010}.  
Conventional grid search methods for learning such parameters are computationally prohibitive, since they require numerically solving the Richards equation for a large number of possible parameter values.  
Optimization algorithms are computationally more efficient but can easily get trapped at local minima due to the pronounced nonlinearity of the Richards equation and  the resulting non-smooth dependence of PDE solutions on its parameters~\citep{Janssen1995,Rasiah1992}. 
Further, real‐world agricultural data are often sparse and noisy, which makes the performance of point estimation methods even more unstable. 
To address these challenges in statistical inference with the Richards equation, we develop in this paper a general Bayesian methodology for learning nonlinear PDE models that can produce more robust parameter estimates with uncertainty quantification.

\subsection{Model and problem formulation} \label{sec:intro.model} 

Consider $n$ observations $(\mathbf{x}_i, \tilde{y}_i)_{i=1}^n$, where $\mathbf{x}_i= ( x_{i,1},x_{i,2},\dots,x_{i,m} )^T\in\mathbb{R}^m$ is the input vector representing time or space in the underlying physical system, and $\tilde{y}_i \in \mathbb{R}$ denotes the noisy response. We model the relationship between $\mathbf{x}_i$ and $\tilde{y}_i$   by 
\begin{equation}
 \tilde{y}_i = f(\mathbf{x}_i ) + e_i,  \text{ for } i = 1, \dots, n,  \label{eq:model}
\end{equation}
where $e_1, \dots, e_n$ are i.i.d. random errors following the normal distribution $N(0,\sigma^2)$, and $f \colon \mathbb{R}^m \rightarrow \mathbb{R}$ is the unknown regression function. 
We assume that $f$ solves the differential equation 
\begin{equation}
\label{eq:implicitpde}
(\mathcal{G}_{\btheta}  f ) (x) = 0,  
\end{equation}  
where $\mathcal{G}$ is a known differential operator parameterized by  the vector $\bm{\theta}=( \theta_1,\dots,\theta_p )^T$ taking values in $\Theta \subset  \mathbb{R}^p$; $\mathcal{G}$ may be nonlinear and inhomogeneous.  
We assume that the boundary and initial value conditions are given such that for each $\btheta \in \Theta$, the PDE~\eqref{eq:implicitpde} admits a unique solution which we denote by $f_{\btheta}$.   When $m = 1$, equation~\eqref{eq:implicitpde} becomes an ordinary differential equation (ODE). 
By equation~\eqref{eq:model}, $\tilde{y}_i$ is a noisy measurement of $y_i \coloneqq f(\mathbf{x}_i)$, which is often called the state variable and cannot be directly observed. 

Our goal is to recover the PDE parameter $\bm{\theta}$ from the observed data, which is also known as the inverse problem with $\mathcal{G}_\theta$ being the forward model. 
From the frequentist viewpoint, we may aim to maximize the log-likelihood of $\bm{\theta}$, 
\begin{equation}\label{eq:likelihood}
\ell(\btheta\,|\,\mathbf{\tilde{y}}, \bX,\sigma^2) 
= -\frac{1}{2\sigma^2}\sum_{i=1}^n[\tilde{y}_i-f_{\btheta}( \mathbf{x}_i ) ]^2, 
\end{equation}
where $\mathbf{\tilde{y}}=(\tilde{y}_1,\dots,\tilde{y}_n)^T\in\mathbb{R}^n$. 
In the Bayesian framework, we put a prior distribution $\pi_0$ on $\bm{\theta}$ and calculate the posterior distribution up to a normalizing constant by the Bayes rule: 
\begin{equation}
\label{eq:posterior}
\pi(\bm{\theta}\,|\,\mathbf{\tilde{y}}, \bX, \sigma^2 )\propto\pi_0(\bm{\theta})e^{\ell(\btheta\,|\,\mathbf{\tilde{y}}, \bX, \sigma^2) },  
\end{equation}
where $\sigma^2$ can be either specified by the user or integrated out by putting an inverse-gamma prior on it. Henceforth we refer to~\eqref{eq:posterior} as the true posterior distribution. Markov chain Monte Carlo (MCMC)  methods can be implemented to sample from  $\pi(\bm{\theta}\,|\,\mathbf{\tilde{y}}, \bX)$. For example, \citet{Gagnon2023} proposed to use locally-balanced multiple-try Metropolis algorithm to sample from the posterior distribution for time-delay differential equation models, since such posterior distributions exhibit irregular behavior and lack closed-form gradients. 
\citet{BESKOS2017} devised several sampling algorithms by leveraging the idea of geometric MCMC to speed up mixing times. More discussion on MCMC-based Bayesian inverse problem can be found in, e.g.,~\citet{Hoang2013, CUI2016}. Variational inference has also been employed to approximate the posterior. \citet{POVALA2022} chose a family of Gaussian distributions parametrized by sparse precision matrices, and \citet{WU2024} utilized three types of normalizing flow to construct the approximate posterior for elliptic PDEs.

However, both maximum likelihood estimation and Bayesian inference methods for the true posterior discussed above require solving the PDE~\eqref{eq:implicitpde}. 
When it does not admit an analytical solution, which is the setting we focus on in this paper,  numerical methods (e.g. finite difference) are needed to compute $f$ and can be very time-consuming when a large number of values of $\bm{\theta}$ are evaluated. Moreover, the posterior distribution is typically multimodal, which presents challenges to both optimization and sampling.

\subsection{Review of existing works}\label{sec:literature}
A number of approaches have been developed in the literature for estimating $\bm{\theta}$ that can significantly reduce the number of calls to the exact numerical solvers. 
The key differences lie in how the PDE constraint is enforced and whether certain structure of the underlying PDE (e.g., linearity) is assumed. 

The first class of methods target the true  posterior distribution defined in equation~\eqref{eq:posterior} or the log-likelihood defined in~\eqref{eq:likelihood} but replace $f_{\btheta}$ by a computationally affordable approximation,  which we denote by $\hat{f}_{\btheta}$.  
This approach includes surrogate modeling techniques~\citep{Zhang2020}, where $\hat{f}$ is treated as a function of $(\mathbf{x}, \bm{\theta})$ and learned by fitting a pre-specified model to the true PDE solution $f_{\btheta}$ evaluated at certain values of $\btheta$. 
Common choices of the surrogate model include multivariate Gaussian process model~\citep{WANG2014167, tang2025gaussian}, polynomial chaos expansions~\citep{LI2015173}, and neural networks~\citep{yan2020}. 
The training  is often performed in an iterative fashion~\citep{Zhang2020}, where new values of  $\btheta$ are drawn from the surrogate-based approximate posterior and the corresponding PDE solutions are computed to update $\hat{f}$.  
Achieving a high-accuracy surrogate model typically requires solving the PDE exactly for a large number of values of $\bm{\theta}$, which can still be computationally too expensive.   
For the application of surrogate modeling in Bayesian variational inference, see, e.g.,~\citet{WU2024}.  
Another approach, which we call ``Gaussian process conditioning,'' completely eliminates the need for a numerical solver by assuming linearity of the operator $\mathcal{G}$. Specifically,  for each fixed $\btheta$,  $\hat{f}_{\btheta}$ is treated  as a random function endowed with a Gaussian process prior distribution, and the posterior distribution of $\hat{f}_{\btheta}$ is  computed by conditioning on $\{ (\mathcal{G}_{\btheta} \hat{f}_{\btheta} )(\bm{x}) = 0 \colon \bm{x} \in \mathcal{X}_0 \}$, where $\mathcal{X}_0$ is a finite set of points chosen by the experimenter (they usually differ from the observed design points.)     
The linearity of  $\mathcal{G}$  preserves Gaussianity, allowing the posterior of $\hat{f}_{\btheta}$ to be expressed in closed form~\citep[Prop. 1]{Cockayne2017}. 
\cite{wang2021} extended this approach to some nonlinear initial value problems with   Dirichlet boundary conditions. They linearized the nonlinear part of the differential operator so that exact Gaussian conditioning can be performed, but this linearization introduces bias to the resulting approximation of $f_{\btheta}$.  Another extension developed by \cite{Li2024} uses optimization algorithms to maximize the un-normalized joint posterior of $(\btheta, \hat{f}_{\btheta})$, which is applicable to nonlinear PDEs reducible to augmented linear forms. 
Note that for both surrogate modeling and Gaussian process conditioning,  the observed data $(\mathbf{x}_i, \tilde{y}_i)_{i = 1}^n$ may not be directly used in the construction of $\hat{f}$. 
 
Instead of directly approximating the PDE solution, an alternative strategy is to assume that the function $f$ belongs to   a large, general-purpose function class parameterized by $\bm{\lambda}$.
The log-likelihood function for $\blambda$ given the data is defined by~\eqref{eq:likelihood} with $\btheta$ replaced by $\blambda$, and  penalty terms are introduced to enforce constraints arising from the PDE~\eqref{eq:implicitpde}. 
Typically, such constraints are constructed by assuming 
\begin{equation}\label{eq:pseudo-like-pde}
    \xi_i \coloneqq ( \mathcal{G}_{\btheta} f  )(\mathbf{x}_i) \overset{\mathrm{i.i.d.}}{\sim}  N(0,  \sigma_{\xi}^2 ), \quad \text{ for }  i = 1, \dots, n. 
\end{equation}
The vector $\bm{\xi} = (\xi_i)_{i=1}^n$ is not  observed, but the resulting likelihood of $(\btheta, \bm{\lambda})$ can be computed by integrating over the distribution of $f_{\bm{\lambda}}$ for each given $\bm{\lambda}$.   
This  can be generalized by allowing $\bm{\xi}$ to follow a multivariate normal distribution with mean zero and a general covariance matrix.  \citet{Xun2013} proposed such a hierarchical Bayesian approach, where $f_{\bm{\lambda}}$ is modeled using pre-specified B-spline basis functions with $\bm{\lambda}$ denoting the coefficient vector. The posterior distribution is computed according to~\eqref{eq:posterior}, with $\btheta$ replaced by $\blambda$, and the prior distribution for $\bm{\lambda}$ given $\btheta$ is  constructed in light of~\eqref{eq:pseudo-like-pde}. 
For ODE models,~\citet{Ramsay2007} proposed a similar   frequentist profiling estimation method where $f_{\bm{\lambda}}$ is also represented by a basis function expansion.  
This idea was extended to settings where the data contain outliers or $\btheta$ itself depends on $\mathbf{x}$~\citep{Cao2011, Cao2012}.  
\citet{Zhou2022} proposed to model $f_{\bm{\lambda}}$ using Gaussian processes, but for nonlinear PDEs, their method requires a linearization step  which needs to be performed on a case-by-case basis and may introduce numerical instability especially when high-order derivatives are involved.

The last class of methods, which we refer to as Gaussian process collocation (GPC), employs a two-stage procedure and estimates $\btheta$ solely based on~\eqref{eq:pseudo-like-pde}, mirroring the collocation principle used in classical PDE solvers.   
This approach was initiated by~\citet{Calderhead2008}, who  considered an ODE of the form $(\mathcal{G}_{\btheta} f)(x) = f'(x) - u(x, f(x), \btheta)$  for some known function $u$. 
The first step of their method is to model $f$ as a Gaussian process with hyperparameter $\bm{\phi}$ and estimate the mean and covariance functions using the observed data  $(x_i, \tilde{y}_i)_{i = 1}^n$ through the standard Gaussian process regression; thus, $f$ is learned without using any information about the underlying PDE.  
Since Gaussian processes are closed under differentiation, once the mean and covariance functions for $f$ is determined,  one also obtains a joint Gaussian distribution of the state variables $\mathbf{y} = (y_i)_{i = 1}^n$  and their corresponding derivatives denoted by  $\mathbf{y}' = (y'_i)_{i = 1}^n$. 
In the second step, the likelihood of $\btheta$ given $(\sigma^2_\xi, \mathbf{y}, \bm{\phi})$ is computed according to~\eqref{eq:pseudo-like-pde}, where $\mathbf{y}'$ is marginalized  over its conditional distribution given $(\mathbf{y}, \bm{\phi})$.  
A Gibbs scheme can be further constructed to sample from the joint posterior distribution of $(\sigma^2_\xi, \mathbf{y}, \bm{\phi}, \btheta)$, which yields the posterior mean estimate for $\btheta$.  
\citet{dondelinger13a} and \citet{barber14} considered the same ODE model and extended the GPC framework by coupling the learning of $(\bm{\phi}, \mathbf{y})$ with that of the underlying ODE, thereby replacing the two-stage procedure with a joint inference scheme. 
A similar method was proposed in~\cite{RAISSI2017} for PDEs where  $\mathcal{G}_{\btheta} f$ depends linearly on the derivatives of $f$; instead of using MCMC sampling to explore the posterior distribution, they simply computed the maximum likelihood estimators for both $\bm{\phi}$ and $\btheta$.   
However, the feasibility of all the aforementioned GPC methods  crucially relies on the assumption on the form of $\mathcal{G}$ which enables closed-form marginalization over $\mathbf{y}'$.  
For nonlinear PDE models, such schemes are usually not applicable. 
\citet{Rai2019} proposed to extend the method of~\citet{Calderhead2008} as follows. First, a Gaussian process model is fitted to the observed data with hyperparameter $\bm{\phi}$ estimated by maximum likelihood. Next, the residual error $\xi_i$ defined in~\eqref{eq:pseudo-like-pde} is evaluated by plugging in the  mean estimates  of $\mathbf{y}$ and its derivatives from the fitted Gaussian process model,  and an optimization algorithm is used to search for the value of $\btheta$ that minimizes the residual sum of squares.  
Although this method is computationally efficient, finding the globally optimal value for $\btheta$ is often challenging due to the multimodality of the objective function. Moreover, the resulting tends to be unstable since the state variables and their derivatives are simply estimated by the means without accounting for the uncertainty.

\subsection{Main contributions}\label{sec:contribute} 
 
In this paper, we propose a new GPC method for estimating unknown parameters in general nonlinear PDE models. 
Our approach has three main components. First, as in the existing GPC methods,  by fitting a Gaussian process model to the observed data $(\bx_i, \tilde{y}_i)_{i=1}^n$, we obtain a joint  distribution of the state variable and its derivatives; denote this distribution by $p( \cdot \, | \, \tilde{\mathbf{y}}, \bX, \hat{\bm{\phi}} )$, where $\hat{\bm{\phi}}$ is an estimator of the hyperparameters of the Gaussian process. 
Second, following~\citet{Calderhead2008}, we define a target posterior distribution for $\bm{\theta}$ by assuming the data-generating model~\eqref{eq:pseudo-like-pde}, which should concentrate around values that approximately satisfy the PDE~\eqref{eq:implicitpde} at design points.  
Due to the nonlinearity of $\mathcal{G}$,  this posterior does not admit a closed form (up to a normalizing constant). We propose to approximate it by drawing $N$ samples from $p( \cdot \, | \, \tilde{\mathbf{y}}, \bX, \hat{\bm{\phi}} )$; denote the resulting posterior   by $\pi^N_{\mathrm{GP}}(\btheta)$.  
This step is key to ensuring the uncertainty quantification, and it also enables us to avoid computing the PDE solution numerically. Note that initial and boundary conditions are not required either. 
Third, since $\pi^N_{\mathrm{GP}}$ differs from the true posterior distribution given by~\eqref{eq:posterior},  we propose a correction procedure where importance sampling with thinned MCMC samples from the approximate posterior $\pi^N_{\mathrm{GP}}$ is used to unbiasedly recover the true posterior. While this step requires access to an exact solver for the PDE~\eqref{eq:implicitpde}, it only needs to be invoked a limited number of times.  
Alternatively, we propose an optimization-based method for utilizing the approximate posterior $\pi^N_{\mathrm{GP}}$. We devise a  prior-guided Bayesian optimization method with the true posterior as the objective function and $\pi^N_{\mathrm{GP}}$ as the ``prior''.  To our knowledge, neither sampling or optimization-based correction procedures has been considered in existing GPC methods. 

We apply our proposed method to statistical inference for the Richards equation with the sink term. Although this problem has been studied extensively in the agricultural literature, most existing approaches rely on computationally intensive techniques such as grid search~\citep{Dong2010,RITTER2003}. Among the few works employing statistical methods that bypass numerical solvers~\citep{Rai2019}, the focus has been limited to the simplest form of the Richards equation without the sink term.  However, as explained earlier, the accurate quantification of the sink term is of significant scientific interest to agricultural researchers, and estimation of the involved parameters often poses a more difficult challenge.  This study represents the first systematic effort to investigate the sink term estimation problem in the Richards equation using collocation methods.  
We obtain accurate and robust estimates across different scenarios in the simulation study, and we apply our method to a real data set on the soil water content measured over three months on a mixed-grass prairie by~\citet{Dong2010}. In that work, the measured soil water content, plus some limited data of soil hydraulic properties, was utilized to model soil water flow and plant root uptake during the growing seasons. However, key root parameters such as maximum rooting depth $L_m$ and root distribution factor $\beta$ could not be directly inferred from domain knowledge or measured environmental covariates. 
We use the proposed GPC method to estimate these parameters, obtaining results consistent with those reported in~\citet{Dong2010} while also providing uncertainty quantification.

\subsection{Overview of the paper}
\label{subsec:overview}  
The rest of this article is organized as follows.  In Section~\ref{sec:prelim}, we review the properties and modeling of Gaussian processes and their derivatives. 
Section~\ref{sec:mainmeth} presents the proposed method for estimating PDE parameters. In Section~\ref{sec:sim}, we describe  the parameter estimation problem for the Richards equation with sink term and conduct simulation studies to evaluate the performance of our approach against existing methods. In Section~\ref{sec:realdata}, we apply the proposed method to the real data set. We conclude the paper with a brief discussion in Section~\ref{sec:conclusion}.

\section{Gaussian process modeling}  \label{sec:prelim} 

Our proposed methodology begins by applying Gaussian process regression to the observed data $(\bx_i, \tilde{y}_i)_{i=1}^n$, which yields a joint distribution of the state variable and its derivatives evaluated at any fixed points. 
For completeness, we provide a brief review of this procedure in this section.  Note that no knowledge about the differential equation~\eqref{eq:implicitpde} or its parameterization $\btheta$ is used in this step. 

\subsection{Notation} We first set and recall some  notation that will be used throughout this paper. 
In general, we use lowercase letters for scalars, bold lowercase for vectors, and bold uppercase for matrices.
Vectors are all assumed to be column vectors.  
Identity matrix is denoted by $\mathbf{I}$. 
The underlying PDE model is given by equation~\eqref{eq:implicitpde} with $\mathbf{x}=( x_1, \dots, x_m )^T\in\mathbb{R}^m$ denoting the input. For the $i$-th observation, we use $\mathbf{x}_i\in\mathbb{R}^m$ to denote its input  and $y_i = f(\mathbf{x}_i)$ to denote the noise-free state variable. 
Let $\mathbf{y}=(y_1,\dots,y_n)^T\in\mathbb{R}^n$, and we denote the  design matrix by  $\mathbf{X}=(\mathbf{x}_1,\dots,\mathbf{x}_n)^T\in\mathbb{R}^{n\times m}$.  
Denote a test data set with $s$ design points by $\mathbf{X_*}=(\mathbf{x}_{*1},\dots,\mathbf{x}_{*s})^T\in\mathbb{R}^{s\times m}$, and denote the corresponding state variable vector by $\mathbf{y_*}\in\mathbb{R}^s$.  
Given a function $m(\,\cdot\,):\mathbb{R}^m\rightarrow\mathbb{R}$, we write $m(\bX) = ( m(\bx_1), \dots, m(\bx_n))^T\in\mathbb{R}^n$. 
Given $k(\,\cdot\,,\,\cdot\,):\mathbb{R}^m\times\mathbb{R}^m\rightarrow\mathbb{R}$, we let $k(\bX, \bX_*)\in\mathbb{R}^{n\times s}$ be the matrix whose $(i, j)$-th entry is given by $k(\mathbf{x}_i,\mathbf{x}_{*j})$. 
When denoting partial derivatives of state variables,  subscript indicates the variable with respect to which differentiation is taken, and superscript indicates   the order of differentiation.  For example, $\partial^1_d \by \in\mathbb{R}^n $  denotes the vector whose $i$-th entry equals $\frac{\partial f}{\partial x_d}\big|_{\bx=\bx_i}$, and similarly, $\partial^{(1, 2)}_{e, d} \by \in\mathbb{R}^n$  denotes the vector whose $i$-th entry equals the mixed partial derivative $\frac{\partial^3 f}{\partial x_e \partial^2 x_d}\big|_{\bx=\bx_i}$.

\subsection{Gaussian process regression} \label{sec:GP} 
Assume that the function $f$ is distributed as a Gaussian process with mean function $m(\,\cdot\,)$ and covariance function $k(\,\cdot\,,\,\cdot\,)$;
denote it by 
\begin{equation}\label{eq:GP.f}
    f(\bx) \sim \mathrm{GP}( m(\bx), k(\bx, \bx')). 
\end{equation}
Throughout this work, we consider the Gaussian kernel 
\begin{equation}
k(\mathbf{x}_i,\mathbf{x}_j)=\sigma_s^2\exp\left\{-\frac{1}{2}\sum^m_{d=1}\frac{(x_{i,d}-x_{j,d})^2}{l_d^2}\right\},
\end{equation}
where $\sigma^2_s$ is the variance parameter and $l^2_1,\dots,l^2_m$ are the length scale parameters.  
We call  the vector $\bm{\phi} \coloneqq \{l_1,\dots,l_m,\sigma_s^2,\sigma^2_y\}$ the Gaussian process hyperparameter. Our methodology can be straightforwardly implemented with other kernels that are sufficiently smooth. 
The distribution of $f$ given by~\eqref{eq:GP.f} implies that  
\begin{equation}\label{eq:GP}
   \mathbf{y} \mid \bX, \bm{\phi} \sim \cN_n (m(\mathbf{X}),\mathbf{K}  ), 
\end{equation}
where $\cN_n$ denotes the $n$-dimensional Gaussian distribution  and $\mathbf{K} = k(\bX, \bX)\in\mathbb{R}^{n\times n}$.  
Since the observation measurement $\mathbf{\tilde{y}} = (\tilde{y}_i)_{i=1}^n$ is considered as a noisy version of $\mathbf{y}$, we can model it as 
\begin{align}
\label{eq:condytil}
\mathbf{\tilde{y}} \mid \bX, \by, \bm{\phi} \sim  \cN_n(\by,\sigma^2_y \mathbf{I}), 
\end{align} 
which yields  
$\mathbf{\tilde{y}} \mid \bX \sim \cN_n(m(\mathbf{X}),\mathbf{K}  +\sigma_y^2\mathbf{I}).$
 
Since the covariance matrix is often powerful enough to capture the trend and variation of $f$, we assume the mean function $m$ is either known or set to the zero function.   
To estimate the hyperparameters $\bm{\phi}$, we take the standard maximum marginal likelihood approach~\citep{GPML2005}.  
The marginal likelihood of $\bm{\phi}$ is calculated as
$$
p(\mathbf{\tilde{y}}\,|\,\mathbf{X},\bm{\phi}) =\int p(\mathbf{\tilde{y}}, \by \,|\,\mathbf{X},\bm{\phi}) \,\dd \mathbf{y} = \int p(\mathbf{\tilde{y}}\,|\,\bX,\by,\bm{\phi})p(\mathbf{y}\,|\,\mathbf{X},\bm{\phi})\,\dd \mathbf{y},
$$
where $p(\mathbf{y}\,|\,\mathbf{X},\bm{\phi})$ denotes the distribution given by \eqref{eq:GP} and $p(\mathbf{\tilde{y}}\,|\,\bX,\by,\bm{\phi})$  denotes that given by \eqref{eq:condytil}. The log-likelihood has the following closed form:
\begin{align}
\label{eq:logmarginlik}
    \log p(\mathbf{\tilde{y}}\,|\,\mathbf{X},\bm{\phi})=-\frac{1}{2}(\mathbf{\tilde{y}}-m(\mathbf{X}))^T\mathbf{K}_y(\mathbf{\tilde{y}}-m(\mathbf{X}))
    -\frac{1}{2}\log|\mathbf{K}_y|-\frac{n}{2}\log2\pi, 
\end{align}
where $\mathbf{K}_y=\mathbf{K}+\sigma_y^2\mathbf{I}$. 
We perform the optimization  with respect to the logarithm of $\bm{\phi}$ for  two reasons. First,   the elements of $\bm{\phi}$ should be positive, and taking logarithm transforms the task to an unconstrained optimization problem. 
Second, parameters in physical systems may have very different scales, and taking logarithm forces the Gaussian process hyperparameters to have similar magnitudes, which help prevent numerical instability. 
Many gradient-based optimization algorithms can be employed to find the optimal value of $\bm{\phi}$ that maximizes the likelihood, and we give the derivatives of \eqref{eq:logmarginlik} with respect to the logarithm of $\bm{\phi}$ in Appendix~\ref{Appen:derimag}.

\subsection{Predictive distribution} \label{sec:gp-predictive}
Let $\bm{\phi}$ be fixed, and consider the predictive distribution of $\by_*$ given a new design matrix $\bX_* \in\mathbb{R}^{s\times m} $.
Let $\mathbf{K}_* = k(\bX,\bX_*)\in\mathbb{R}^{n\times s}$ be the covariance matrix evaluated at all pairs of design points in training and test data, and let  $\mathbf{K}_{**}=k(\bX_*,\bX_*)\in\mathbb{R}^{s\times s}$ be the kernel of the test data. 
The predictive distribution of  $\mathbf{y}_*$ is given by  
\begin{align}
\label{eq:preddist}
 \mathbf{y}_* \mid \mathbf{\tilde{y}},\mathbf{X},\mathbf{X}_*, \bm{\phi} \sim \cN_s( \bm{\mu}_*, \mathbf{\Sigma}_*),
\end{align}
where 
\begin{align*}
\bm{\mu}_*&=m(\mathbf{X_*})+\mathbf{K}^{T}_*(\mathbf{K}+\sigma_y^2\mathbf{I})^{-1}(\mathbf{\tilde{y}}-m(\mathbf{X})), \\ 
\mathbf{\Sigma}_*&=\mathbf{K}_{**}-\mathbf{K}^{T}_*(\mathbf{K}+\sigma_y^2\mathbf{I})^{-1}\mathbf{K}_*.
\end{align*}
We can think of \eqref{eq:preddist} as the posterior distribution of the state variable evaluated at $\mathbf{X_*}$ given the observed data, which has mean $\mu_*$ and covariance matrix $\Sigma_*$. 
In particular, this fitted Gaussian process model also yields the predictive distribution of the unobserved state variable vector $\mathbf{y}$ and its derivatives (see the next subsection), which we denote by $p( \cdot \mid \mathbf{\tilde{y}},\mathbf{X}, \bm{\phi} )$.

\subsection{Gaussian process derivatives}
\label{sec:gp-deri}
Since differentiation is a linear operator, the derivative of a Gaussian process remains a Gaussian process~\citep{bogachev1998gaussian, solak2002}. Moreover,  state variables and their partial derivatives  can be modeled as a joint Gaussian distribution. 
Assuming that $m(x)$ is a constant function, the mean function of the derivatives of $f$ is zero everywhere.  The covariance function between the state variables and the $p$-th order derivatives with respect to the $d$-th coordinate is given by 
\begin{align*}
      &\text{cov}\left(y_i, \; \frac{\partial^p f}{\partial x_d}\big|_{\bx=\bx_j}\right)=\frac{\partial^p k(\mathbf{x}_i,\mathbf{x}_j)}{\partial x_{j,d}^p}, \quad \text{ for } i, j = 1, \dots, n, 
\end{align*}
and similarly, the covariance function between the state variable derivatives is given by  
\begin{align*} 
\text{cov}\left(\frac{\partial^q f}{\partial x_e}\big|_{\bx=\bx_i}, \; \frac{\partial^p f}{\partial x_d}\big|_{\bx=\bx_j}\right)=\frac{\partial^{p+q} k(\mathbf{x}_i,\mathbf{x}_j)}{\partial x_{i,e}^q\partial x_{j,d}^p}, 
\quad \text{ for } i, j = 1, \dots, n. 
\end{align*}
To simplify the notation, we use the following matrix notation: 
\begin{equation}
   \partial^{(q, p)}_{e, d} \mathbf{K} = \left[\frac{\partial^{q+p} k(\mathbf{x}_i,\mathbf{x}_j)}{\partial x_{i,e}^q\partial x_{j,d}^p}
   \right]_{i, j = 1}^n; 
\end{equation}
when $q = 0$ or $p = 0$, we write 
\begin{equation}
    \partial^{(q, 0)}_{e} \mathbf{K} = \left[ \frac{\partial^{q} k(\mathbf{x}_i,\mathbf{x}_j)}{\partial x_{i,e}^q }  \right]_{i, j = 1}^n, \quad 
        \partial^{(0, p)}_{d} \mathbf{K} = \left[ \frac{\partial^{p} k(\mathbf{x}_i,\mathbf{x}_j)}{\partial x_{j,d}^q }  \right]_{i, j = 1}^n. 
\end{equation} 
 
As an example, here we give the predictive distribution of  $(\mathbf{y},\partial^1_1\by,\partial^1_2\by,\partial^2_1\by)$, which will be used later in our analysis of the Richards equation.    Applying the formulas given in the previous subsection, we get 
\begin{equation}\label{eq:fullcond}
\mathbf{y}, \, \partial^1_1\by, \, \partial^1_2\by, \, \partial^2_1\by\,|\,\mathbf{\tilde{y}}, \, \mathbf{X}, \, \bm{\phi} \sim\cN_{4n}(\bm{\mu},\mathbf{\Sigma}),
\end{equation}
where 
\begin{align}
\bm{\mu}&=\begin{bmatrix}
    m(\mathbf{X}),\bm{0},\bm{0},\bm{0}
\end{bmatrix}^T+\bC_1^T(\mathbf{K}+\sigma_y^2\mathbf{I})^{-1}(\mathbf{\tilde{y}}-m(\mathbf{X})),  \label{eq:condmean} \\
\mathbf{\Sigma}&=\bC_2-\bC_1^T(\mathbf{K}+\sigma_y^2\mathbf{I})^{-1}\bC_1. 
\end{align} 
In the above expressions, $\bC_1$ encodes the covariance between $\tilde{\by}$ and $(\mathbf{y},\partial^1_1\by,\partial^1_2\by,\partial^2_1\by)$ and is given by 
\begin{align*}
\bC_1=\begin{bmatrix}\mathbf{K}&\quad\partial^{(0, 1)}_{1} \mathbf{K}&\quad\partial^{(0, 1)}_{2} \mathbf{K}&\quad\partial^{(0, 2)}_{1} \mathbf{K} \end{bmatrix}\in\mathbb{R}^{n\times 4n}, 
\end{align*}
and $\bC_2$ is the covariance matrix of $(\mathbf{y},\partial^1_1\by,\partial^1_2\by,\partial^2_1\by)$, which is given by
\begin{align*}
    \bC_2=
    \begin{bmatrix}
\bK&\quad\partial^{(0, 1)}_{1} \mathbf{K}  
&\quad\partial^{(0, 1)}_{2} \mathbf{K}&\quad\partial^{(0, 2)}_{1} \mathbf{K}\\
\partial^{(1,0)}_{1} \mathbf{K} & \quad\partial^{(1, 1)}_{1, 1} \mathbf{K}&\quad\partial^{(1, 1)}_{1, 2} \mathbf{K}&\quad\partial^{(1, 2)}_{1, 1} \mathbf{K}\\
\partial^{(1, 0)}_{2} \mathbf{K}&  \quad\partial^{(1, 1)}_{2, 1} \mathbf{K}&\quad\partial^{(1, 1)}_{2, 2} \mathbf{K}&\quad\partial^{(1, 2)}_{2, 1} \mathbf{K}\\
\partial^{(2, 0)}_{1} \mathbf{K}&
\quad\partial^{(2, 1)}_{1, 1} \mathbf{K}&
\quad\partial^{(2, 1)}_{1, 2} \mathbf{K}&\quad\partial^{(2, 2)}_{1, 1} \mathbf{K}  \\
\end{bmatrix}\in\mathbb{R}^{4n\times 4n}.
\end{align*}

\section{A new Gaussian process collocation method}
\label{sec:mainmeth} 
In this section, we describe our method for Bayesian statistical inference with general PDE models. Similar to~\citet{Calderhead2008}, we use the collocation principle to construct an computationally efficient posterior distribution which can be approximated by sampling methods. However, the approximate posterior may differ from the true posterior  in  regions where the derivative estimates of $f$ from the fitted Gaussian process are inaccurate. Assuming a numerical solver is available, we propose two methods for correcting for the discrepancy between the approximate posterior and true posterior distributions.  
The first method uses thinned MCMC samples to perform importance sampling correction, and the second employs a  prior-guided Bayesian optimization technique.  

\subsection{Gaussian process collocation posterior}
\label{sec:pdeest}
The true posterior distribution given by~\eqref{eq:posterior} is typically impractical to evaluate due to the high computational cost for numerically solving the PDE~\eqref{eq:implicitpde}. 
Therefore, we construct an alternative posterior distribution based on~\eqref{eq:implicitpde}.  
We use $\bm{\omega}_i$ to denote the vector obtained by stacking the state variable $\mathbf{y}_i = f(\bx_i)$ and its partial derivatives of all orders involved in the operator $\mathcal{G}$, and we let $G$ be the function such that 
\begin{equation}\label{eq:operator-to-function-G}
    (\mathcal{G}_\theta f)(\bx ) = G( \bx,  \bm{\omega}; \bm{\theta} ). 
\end{equation}
Although $\bm{\Omega} = (\bm{\omega}_1, \dots, \bm{\omega}_n)$ cannot be observed,  we can model its distribution by $p( \bm{\Omega} \mid \mathbf{\tilde{y}}, \bX, \hat{\bm{\phi}}  )$, where $p(\cdot \mid \mathbf{\tilde{y}}, \bX, \hat{\bm{\phi}})$ is the Gaussian process predictive distribution described in Section~\ref{sec:gp-deri}, where $\hat{\bm{\phi}}$ is the maximum likelihood estimator  described in Section~\ref{sec:GP}.  

Recall that the main idea of GPC methods is to construct the likelihood function based on the generative model~\eqref{eq:pseudo-like-pde}, where the residual error can now be expressed by $\xi_i = G( \bx_i,  \bm{\omega}_i; \bm{\theta} )$. 
We select $n_s$ points from the observed design matrix $\bX$ and denote them by $\bx_{(1)}, \dots, \bx_{(n_s)}$. 
Define 
\begin{equation}\label{eq:xi-vec}
    \bm{\xi}(\bm{\theta}) \coloneqq \{G(\mathbf{x}_{(1)},\bm{\omega}_{(1)};\bm{\theta}), \, \dots, \,  G (\mathbf{x}_{(n_s)},\bm{\omega}_{(n_s)};\bm{\theta})\}^T\in\mathbb{R}^{n_s}, 
\end{equation}
which is the residual vector at the selected points. 
The notation $\bm{\xi}(\bm{\theta})$ emphasizes the dependency on $\bm{\theta}$, 
but note that given $\btheta$, it is a random element whose distribution can be determined from the Gaussian process predictive distribution for $\bm{\omega}_{(1)}, \dots, \bm{\omega}_{(n_s)}$. 
We define the GPC posterior distribution of $\bm{\theta}$ by  
\begin{equation}\label{eq:gpposterior}
\pi_{\rm{GP}}(\bm{\theta}\,|\, \mathbf{\tilde{y}}, \bX)\propto \pi_0(\bm{\theta})\mathbb{E}\left\{\exp\left[-\frac{1}{2}\bm{\xi}(\bm{\theta})^T\bm{\Sigma}_{n_s}^{-1}\bm{\xi}(\bm{\theta})\right]\right\},
\end{equation}
where $\pi_0(\bm{\theta})$ is the prior distribution, and the expectation is taken with respect to the predictive distribution $p(  \bm{\omega}_{(1)}, \dots, \bm{\omega}_{(n_s)}  \mid \mathbf{\tilde{y}}, \bX, \hat{\bm{\phi}}  )$.  
That is, unlike~\eqref{eq:pseudo-like-pde}, we derive the likelihood by pretending that, when $\btheta$ takes the true value,  $\bm{\xi}$ follows a zero-mean multivariate normal distribution with a covariance matrix $\bm{\Sigma}_{n_s}$.  
Expert knowledge can be encoded into $\pi_0(\bm{\theta})$, and when such knowledge is not available, we can use a noninformative prior distribution (e.g. normal prior with large variance). 
Regarding the choice of  $\bm{\Sigma}_{n_s}$, we employ an empirical method, where we simulate the residual vector $\bm{\xi}(\bm{\theta})$ using a reasonable guess for $\bm{\theta}_0$ (which, for example, can be obtained from a preliminary experiment), and estimate $\bm{\Sigma}_{n_s}$ using the sample covariance matrix.

For nonlinear PDEs, the expectation in equation~\eqref{eq:gpposterior} cannot be evaluated in closed form, and thus we further approximate it using a sampling-based strategy. 
We draw $N$ samples of the vector $(\bm{\omega}_{(1)},\dots,\bm{\omega}_{(n_s)})$ from $p(  \cdot \mid \mathbf{\tilde{y}}, \bX, \hat{\bm{\phi}}  )$ and denote the resulting residual vector by $\bm{\xi}^{(1)}(\bm{\theta}), \dots, \bm{\xi}^{(N)}(\bm{\theta})$. Then, we approximate the expectation in \eqref{eq:gpposterior} using an average of these samples, which yields the approximate posterior distribution: 
\begin{align}\label{eq:gpposteriorave}
\pi_{\rm{GP}}^N(\bm{\theta}\,|\,\mathbf{\tilde{y}}, \bX)\propto \pi_0(\bm{\theta})\sum_{i=1}^N\exp\left\{-\frac{1}{2}\bm{\xi}^{(i)}(\bm{\theta})^T\bm{\Sigma}_{n_s}^{-1}\bm{\xi}^{(i)}(\bm{\theta})\right\}.
\end{align}
The posterior distribution $\pi_{\rm{GP}}^N$ is not analytically tractable, but it is straightforward to implement MCMC sampling methods targeting it. When the dimension of $\bm{\theta}$ is low, one can simply use the standard Metropolis--Hastings algorithm~\citep{Hastings1970}. 
Alternatively, one can also construct a Metropolis-within-Gibbs scheme which sequentially updates each entry of $\btheta$  via either a Metropolis--Hastings or Gibbs step.  
When $\mathcal{G}$ is highly nonlinear, $\pi_{\rm{GP}}$ can often be multimodal, and a temperature parameter could be further introduced to flatten the posterior landscape, which can improve the MCMC sampling efficiency and make it easier to detect multiple modes. 
Though not considered in this work, this method can be further extended by letting the $n_s$ design points be randomly generated (which introduces an additional expectation in~\eqref{eq:gpposterior} over this distribution) and resampling the design points when generating $\bm{\xi}^{(1)}(\bm{\theta}), \dots, \bm{\xi}^{(N)}(\bm{\theta})$. 

Compared to existing methods such as the GPC scheme for nonlinear PDE models proposed by~\citet{Rai2019}, our method introduces several key distinctions. 
First, we define the target GPC posterior by averaging over samples of state variables and their derivatives, in contrast to the point estimate approach taken by~\citet{Rai2019}. This averaging is crucial for capturing the uncertainty in the predictive distribution  $p( \cdot  \mid \mathbf{\tilde{y}}, \bX, \hat{\bm{\phi}}  )$. For nonlinear PDE models, derivative estimates can exhibit  high variance and be very unstable near the boundaries. Hence, if one simply plugs in the mean estimate for $(\bm{\omega}_{(1)},\dots,\bm{\omega}_{(n_s)})$, the resulting estimate for $\bm{\theta}$ can be highly biased and sensitive to certain data points. 
Second, we propose to construct the GPC posterior using only a subset of observations instead of all of them. This step not only alleviates the computational burden but  also allows us to deliberately exclude points near the boundary in order to mitigate the adverse effects of poor derivative estimates in those regions from the fitted Gaussian process model. 
Third,  we estimate the covariance matrix of $\bm{\xi}$ rather than assuming its entries are i.i.d., which also helps ensure that the estimates for $\bm{\theta}$ are not disproportionately affected by a small number of  data points where the residual exhibits large variance.

\subsection{Importance sampling correction}
\label{subsec:IS}
The methodology proposed in the previous subsection enables the estimation of $\btheta$ without using a numerical solver for the PDE~\eqref{eq:implicitpde}. However, it is entirely possible that the landscape of $\pi_{\rm{GP}}^N$ and that of the true posterior~\eqref{eq:posterior} differ significantly in certain regions. 
To address this, we propose to use importance sampling (IS) to correct for the discrepancy. Although this method requires access to the numerical solver, we only need to invoke it a small number of times, and thus the overall runtime remains low compared to methods such as surrogate modeling. 
 
The true posterior~\eqref{eq:posterior} depends on the unknown error variance $\sigma^2$, but we can integrate it out by assuming an inverse-Gamma distribution with parameters $(\alpha, \eta)$, which yields 
\begin{align}
     \label{eq:exactpost}
     \pi(\bm{\theta}\,|\,\mathbf{\tilde{y}}, \bX )
&\propto \pi_0(\bm{\theta})\left\{\sum_{i=1}^n\frac{[\tilde{y}_i-f_{\btheta}(\bx_i)]^2}{2}+\eta\right\}^{-\alpha-\frac{n}{2}}.
\end{align}
Let $\bm{\theta}^{\rm{imp}}_1, \dots, \bm{\theta}^{\rm{imp}}_{n_t}$ denote $n_t$ samples generated from the MCMC sampler targeting $\pi_{\rm{GP}}^N$. Note that these samples are typically selected by further thinning the MCMC trajectory, and thus $n_t$ can be much smaller than the number of MCMC iterations. In our experiments, we always use $n_t \leq 15$.

Suppose we aim to learn $\pi(g) \coloneqq \mathbb{E}_{\pi}g(\bm{\theta})$ for some function $g \colon \mathbb{R}^p \rightarrow \mathbb{R}$.
For example, when $g(\bm{\theta})=\bm{\theta}$, $\pi(g)$ is the posterior mean estimate for $\btheta$.  
We compute the self-normalized importance sampling estimator
\begin{align}
\label{eq:impest} 
\hat{\pi}(g) = \frac{\sum_{i=1}^{n_t}g(\bm{\theta}^{\rm{imp}}_i)\rho(\bm{\theta}^{\rm{imp}}_i)}{\sum_{i=1}^{n_t}\rho(\bm{\theta}^{\rm{imp}}_i)}, \text{ where } \rho(\bm{\theta}^{\rm{imp}}_i)=\frac{\pi(\bm{\theta}_i^{\rm{imp}}\,|\,\mathbf{\tilde{y}}, \bX)}{\pi_{\rm{GP}}^N(\bm{\theta}_i^{\rm{imp}}\,|\,\mathbf{\tilde{y}}, \bX)}. 
\end{align}
Note that the importance weight $\rho(\bm{\theta}^{\rm{imp}}_i)$ only needs to be evaluated up to a normalizing constant. 
When $\btheta$ is one-dimensional,  the highest posterior density (HPD) region of the true posterior can be estimated by considering $g(\theta)=\mathbf{1}(\theta\leq b)$, where  $\mathbf{1}(\cdot)$ is the indicator function.   
When $\btheta$ is multidimensional, we first approximate the true posterior by employing a kernel density estimation  method based on the samples and their weights. For any $\bm{\theta}\in\Theta$, the posterior density is estimated by
\begin{equation}
\label{eq:kde}
    \tilde{\pi}(\btheta) = \sum_{i=1}^{n_t}\bar{\rho}_iK_H(\btheta-\btheta^{\rm{imp}}_i),
\end{equation}
where $\bar{\rho}_i=\rho(\bm{\theta}^{\rm{imp}}_i)/\sum_{i=1}^{n_t}\rho(\bm{\theta}^{\rm{imp}}_i)$ denotes the normalized importance weight and $K_H$ is a Gaussian kernel with bandwidth matrix $H$. The HPD region can then be constructed by using $\tilde{\pi}$. 
We summarize the proposed Gaussian process collocation method with importance correction in Algorithm~\ref{Algo1} and refer to it as GPC-I henceforth. 

\begin{algorithm}
\caption{Gaussian Process Collocation with Importance Correction (GPC-I)}
\label{Algo1}
\begin{algorithmic}[1]
\Require  $\mathcal{G}_{\btheta}$ and its numerical solver, 
 $(\tilde{\mathbf{y}}, \mathbf{X})$,   $\pi_0(\btheta)$,  $\btheta_0$,   
  $n_s$, $n_t$, $N$,  an MCMC sampler on $\Theta$. 
\State Fit a Gaussian process predictive model $p( \cdot  \mid \mathbf{\tilde{y}}, \bX, \hat{\bm{\phi}}  )$ to the observed data  
  $(\tilde{\mathbf{y}}, \mathbf{X})$.   
\State Select $n_s$ design points $\bx_{(1)}, \dots, \bx_{(n_s)}$. 
\State Draw $N$ copies of $(\bm{\omega}_{(1)},\dots,\bm{\omega}_{(n_s)})$  from the predictive distribution $p( \cdot  \mid \mathbf{\tilde{y}}, \bX, \hat{\bm{\phi}} )$.    
\State Set $\bm{\Sigma}_{n_s}$ to the sample covariance matrix of $ \bm{\xi}^{(1)}(\bm{\theta}_0), \dots, \bm{\xi}^{(N)}(\bm{\theta}_0)$.   
\State Generate  thinned post-burn-in MCMC samples $\bm{\theta}_1^{\rm{imp}}, \dots, \bm{\theta}_{n_t}^{\rm{imp}}$
from  $\pi_{\rm{GP}}^N(\bm{\theta}\,|\,\mathbf{\tilde{y}}, \bX)$. 
\State Compute $\pi(\bm{\theta}_i^{\rm{imp}}\,|\,\mathbf{\tilde{y}}, \bX)$ using the numerical solver for $i =1, \dots, n_t$. 
\State Compute the importance sampling estimator~\eqref{eq:impest}.  
\end{algorithmic}
\end{algorithm}

\subsection{Bayesian optimization with a prior for the optimum}
\label{sec:prbo}
Alternatively, we can use optimization methods to search for the optimal parameter value  that maximizes the true posterior \eqref{eq:exactpost}. In our setting, we define the objective function as the negative logarithm of the true posterior.   
Since the true posterior requires solving the PDE~\eqref{eq:implicitpde}, naive optimization methods that require evaluating the objective function at many points can be computationally prohibitive, and gradient-based techniques are not applicable either. 
This motivates us to employ Bayesian optimization~\citep{frazier2018}, which builds a surrogate for the objective function via Gaussian process regression and then selects the next point to evaluate by maximizing a so-called ``acquisition function.'' We denote by $\bm{\Theta}^{(0)}=(\btheta^{(0)}_1,\dots,\btheta^{(0)}_{n_0})$ the set of initial points used to fit a Gaussian process  $U(\cdot)$ at the beginning. 
We use the expected improvement (EI), a widely used acquisition function, to update the set $\bm{\Theta}^{(0)}$. 
In the $t$-th iteration,  EI is   computed as  
\begin{align}
\label{eq:EI}
    \text{EI}_t(\mathbf{\bm{\theta}})=\mathbb{E}[(\min(\mathbf{u})-U(\bm{\theta}))^+\mid U(\bm{\Theta}^{(t - 1)})=\mathbf{u}^{(t - 1)}
    ],
\end{align}
where $\mathbf{u}^{(t)}=(u^{(t)}_1,\dots,u^{(t)}_{n_{t}})$ with $u_i^{(t)}=-\log\pi(\bm{\theta}^{(t)}_i\,|\,\mathbf{\tilde{y}}, \bX )$ and $n_t = n_0 + t$. 
Let $\bm{\theta}_t = \argmax   \text{EI}_t (\btheta) $ denote the value that maximizes EI, and define $u_t = -\log\pi(\bm{\theta}_t \,|\,\mathbf{\tilde{y}}, \bX )$.  
Then, we set $\bm{\Theta}^{(t)} = ( \bm{\Theta}^{(t-1)}, \, \btheta_t)$ and $\mathbf{u}^{(t)}= (\mathbf{u}^{(t-1)}, \, u_t)$, which will be used in the next update. After $B$ iterations,  the procedure terminates by reporting the $\bm{\theta}$ with the minimum objective value across the evaluated set.
 
 We leverage the approximate posterior \eqref{eq:gpposteriorave} to improve the efficiency of the optimization procedure by adopting the Bayesian optimization with a prior for the optimum (BOPrO) method proposed by~\citet{souza2021}. In their approach, EI is calculated based on a pseudo-posterior of $\btheta$ defined by combining the prior with a function $\mathcal{M}(\btheta)$ known as the probability of improvement~ \citep{kushner1964}.
 In our context, the approximate posterior $\pi_{\mathrm{GP}}^N$ is considered as the ``prior'' knowledge indicating regions where the optimal point is likely to lie in. 
 To construct the pseudo-posterior, we 
 first normalize $\pi_{\mathrm{GP}}^N$  to $[0,1]$. 
 Specifically,
  we evaluate $\pi_{\mathrm{GP}}^N$ at a set of $\btheta$'s (e.g., those generated from the MCMC sampling) and find the empirical maximum and minimum values. Standard min–max scaling is then applied, with values greater than 1 clipped to 1 and values less than 0 clipped to 0. 
 Then, we define the pseudo-posterior in the $t$-th iteration as 
\begin{align}
\label{eq:pseudo-post}
p_t(\bm{\theta}\mid u)\propto\begin{cases}
    g_t(\bm{\theta}):=\pi^N_{\rm{GP}}(\bm{\theta})\mathcal{M}_t(\bm{\theta})^{\frac{t}{\tau}}, &\text{if}\quad u<f_{\delta}\\
     b_t(\bm{\theta}):=(1-\pi^N_{\rm{GP}}(\bm{\theta}))(1-\mathcal{M}_t(\bm{\theta}))^{\frac{t}{\tau}}, &\text{if}\quad u\geq f_{\delta}
\end{cases},
\end{align}
where $\mathcal{M}_t(\bm{\theta}) \coloneqq p\left(U(\bm{\theta})<f_{\delta}\mid U(\bm{\Theta}^{(t - 1)})=\mathbf{u}^{(t-1)}\right)$  admits a closed form, $f_{\delta}$ is the  $\delta$-quantile of  $\mathbf{u}$,  
and $\tau$ controls the number of iterations it needs for the function $\mathcal{M}$ to dominate the effect of the prior knowledge $\pi_{\mathrm{GP}}^N$. The EI is calculated as 
$$\text{EI}_t(\bm{\theta})=\int\max(f_{\delta}-u,0) p_t (u\mid\bm{\theta}) \dd u.$$ 
It can then be shown that maximizing $\text{EI}_t(\btheta)$ is equivalent to minimizing $b_t(\bm{\theta})/g_t(\bm{\theta})$. 
The major advantage of BOPrO is that it can accelerate the convergence when the prior is informative, 
while it can also correct for misleading priors through further iterations. 
We call this method BO-GPC and summarize it in Algorithm \ref{Algo2}.

\begin{algorithm}
\caption{Bayesian Optimization guided by Gaussian Process Collocation (BO-GPC)}
\label{Algo2}
\begin{algorithmic}[1]
\Require $\mathcal{G}_{\btheta}$ and its numerical solver, 
 $(\tilde{\mathbf{y}}, \mathbf{X})$,   $\pi_0(\btheta)$,  $\btheta_0$,   
  $n_s$, $N$, $n_0$, $B$, $\delta$, $\tau$.
\State Define the approximate posterior of $\bm{\theta}$ as described in lines~1--4 of Algorithm~\ref{Algo1} and normalize it to [0,1].
\State Randomly select some initial points $\bm{\Theta}^{(0)}$ and compute $\mathbf{u}^{(0)}$.
\For{$t = 1, 2, \dots, B$}
    \State Fit a Gaussian process over $\bm{\Theta}^{(t-1)}$ and $\mathbf{u}^{(t-1)}$.
    \State Let $\bm{\theta}_t = \argmin_{\bm{\theta}}b_t(\bm{\theta})/g_t(\bm{\theta})$ and $u_t=-\log\pi(\bm{\theta}_t\,|\,\mathbf{\tilde{y}}, \bX )$.
    \State Let $\bm{\Theta}^{(t)} = ( \bm{\Theta}^{(t-1)}, \, \btheta_t)$ and $\mathbf{u}^{(t)}= (\mathbf{u}^{(t-1)}, \, u_t)$. 
\EndFor
\State Return $\hat{\bm{\theta}} = \argmax_{\btheta\in\bm{\Theta}^{(B)}}\log\pi(\bm{\theta}\,|\,\mathbf{\tilde{y}}, \bX )$.  
\end{algorithmic}
\end{algorithm}

\section{Bayesian inference for the Richards equation}
\label{sec:sim} 

In this section, we consider the application of our proposed methods to parameter estimation for the Richards equation. We compare Algorithms~\ref{Algo1} and~\ref{Algo2} with two other methods: a simple Bayesian optimization scheme, and the Gaussian process collocation method of \cite{Rai2019}. 

\subsection{Richards equation with the sink term}
The Richards equation models the spatio-temporal dynamics of soil water flow, and an extra sink term $S$ can be added to model the water taken by plants. 
The equation can be written as 
\begin{equation}
\label{eq:richards}
    \frac{\partial f}{\partial t} = \frac{\partial}{\partial z}\left[k\frac{\partial h}{\partial z}\right]+\frac{\partial k}{\partial z}-S(h,z,t;\bm{\theta}), 
\end{equation} 
where $f, h, k$ are all functions of $(z, t)$. The variable $t$ denotes time  in seconds (s) and $z$ denotes the depth in meters (m; positive upwards).  
The function $f(z, t)$ denotes the volumetric soil water content, which is defined as the ratio of water volume to total soil volume. $h(z, t) $ is the soil matric potential (m), and $k(z, t)$ is the hydraulic conductivity ($\text{m}\cdot\text{s}^{-1}$). 
The function $S$  models the water taken by the plant root and is parameterized by $\bm{\theta}$ that we aim to estimate. 
We assume $f > 0, k > 0, h < 0$ and their relationships are determined by the van  Genuchten model~\citep{van1980}:  
\begin{align}
  &f= c_r+(c_s-c_r)\left[1+|\gamma h|^{1/(1-\nu)}\right]^{-\nu}, \label{eq:vanGa} \\
&k=k_{\rm{sat}}C^{1/2}\left[1-\left(1-C^{1/\nu}\right)^{\nu}\right]^2\label{eq:vanGb}, \\  
   &C=\frac{f-c_r}{c_s-c_r}\label{eq:vanGc},
\end{align}  
where $C$ is known as the effective saturation, $k_{\mathrm{sat}}$ ($\text{m}\cdot \text{s}^{-1}$) is the saturated hydraulic conductivity,   $c_r, c_s \in [0, 1]$ are residual and saturated water content respectively, and $\gamma$ ($\text{m}^{-1}$) and $m$ are positive empirical shape parameters. 
We assume $k_{\mathrm{sat}}$, $c_r$, $c_s$, $\gamma$ and $\nu$ are known, but they may vary across different soil depth intervals. In the agricultural literature, these hydraulic parameters are either measured directly or predicted using well-established models. 
According to the van Genuchten model, both $k$ and $h$ can be written as some nonlinear functions of $f$, and thus the Richards equation~\eqref{eq:richards} can be rewritten as $(\mathcal{G}_\theta f)(z, t) = 0$ for some nonlinear differential operator $\mathcal{G}_\theta$. See Appendix~\ref{Appen:ricfact} for the details. Equivalently, as in equation~\eqref{eq:operator-to-function-G}, we can also express this PDE model by $  G( (z, t),  \bm{\omega}; \bm{\theta} ) = 0$,   where 
\begin{equation}
    \bm{\omega} = \left( f(z, t), \, \frac{\partial f}{\partial z}(z, t), \, \frac{\partial f}{ \partial t}(z, t),  \, \frac{\partial^2 f}{\partial z^2}(z, t)  \right)
\end{equation}
denotes the state variable and its derivatives involved in $\mathcal{G}_{\btheta}$. The Gaussian process predictive distribution for $\bm{\omega}$ is given by \eqref{eq:fullcond}.
 
Several root water uptake models have been proposed for representing the sink term $S$ in the agricultural literature; see, e.g., \cite{chandra1996,Yadav2008,LI2001189}. In this study, we adopt the following model considered in~\citet{Dong2010}: 
\begin{equation}
     S(h, z, t;\bm{\theta}) = \left\{\begin{array}{cc}
      0,    & \quad \text{ if }  z > L_m Z_r(t), \\
      \frac{(1 + \beta) \psi(z, t)  }{L_m Z_r(t)}\left( 1 - \frac{|z|}{L_m Z_r(t)}\right)^\beta,   &  \quad \text{ if }  z \leq L_m Z_r(t), 
      \end{array}  
     \right.  
\end{equation}
where $\btheta = (\beta, L_m)^T$ is the unknown parameter vector, $Z_r(t)$ and $\psi(z, t)$ are both known functions that can be computed using covariate information such as the the leaf area index; for details, see Appendix~\ref{Appen:sink}. 
The parameter $L_m$ characterizes the maximum rooting depth of the plant and $\beta$ represents the shape of root distribution within the root zone soil profile. Both parameters  strongly influence the water use efficiency of the plant, but are typically difficult to measure or estimate through explicit formulas. Despite the efforts of numerous researchers, the root distribution shape parameter $\beta$ has been shown to vary greatly for different plant species~\citep{chandra1996}, and even for the same species (such as wheat) under different soil/management conditions~\citep{WuJQ1999,ZuoQ2004}.  We aim to make inference on $\bm{\theta}$ using the proposed GPC methodology. To our knowledge, this is the first work that uses a statistical method to estimate the parameters in the sink term of the Richards equation.   

\begin{figure}
\centering
\includegraphics[width=13cm]{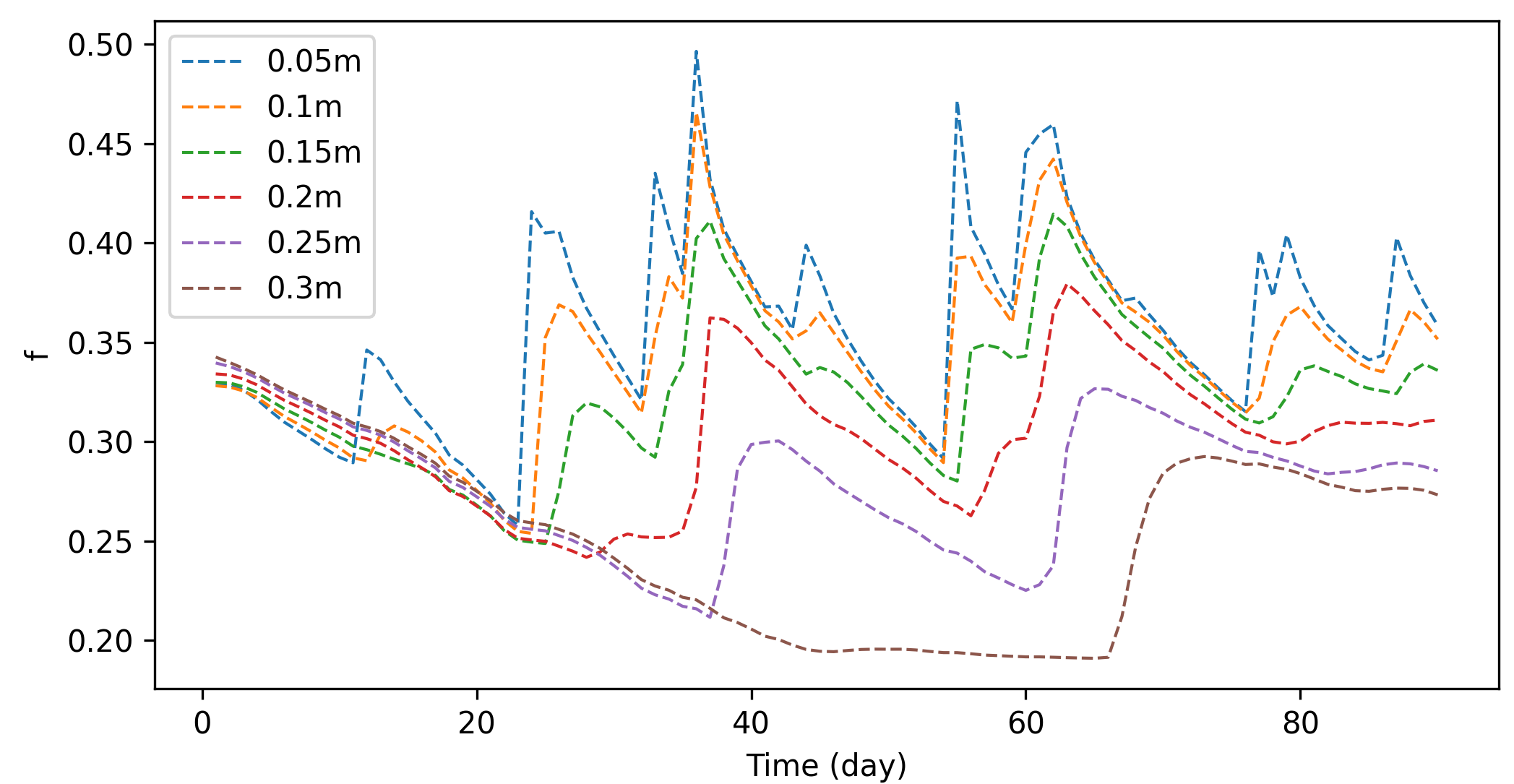}
\caption{Numerical solutions of the soil water content ($L_m=1.4$ and $\beta = 1.9$) across time (day) at depth 0.05m, 0.1m, 0.15m, 0.2m, 0.25m, 0.3m.}
\label{fig:f_sim}
\end{figure}

\subsection{Generation of simulated data}
Since the Richards equation \eqref{eq:richards} is highly nonlinear, numerical solutions are required for the generation of simulated data.  We implement in \texttt{RCpp} the method of~\cite{Celia1990} to simulate the data, which is a finite difference scheme coupled with a mixed-form Picard iteration. The details are provided in Appendix~\ref{Appen:picard}.  This method is fast and mass-conservative, which is essential to the estimation of the sink term. 
We consider a time horizon of 90 days, expressed in seconds, so that the temporal domain is  $t \in [0, \; 90\times 86400]$ (each day has $86400$ seconds). Depth is measured in meters, and the spatial domain is given by $z \in [-0.3, \, -0.01]$.  
For the finite difference scheme, time and depth are discretized with step sizes $\Delta t=400$ and $\Delta z =0.005$, respectively. Let $(z_1=-0.3,z_2,\dots,z_M=-0.01)$ be the negatives of depth grid points.

In our simulation study, the soil parameters appearing in the van Genuchten model are assumed to be homogeneous across depth  and are specified as $c_r = 0.156$, $c_s=0.60$, $\gamma=5.87$, $m=0.273$ and $k_{\mathrm{sat}}=6\times10^{-6}$. 
Three initial and boundary conditions are needed for the PDE solver. The initial condition is set as $f(z,t=0)=0.33+0.5\times(z+0.1)^2$. 
Following~\citet{Dong2010}, we let the upper boundary condition be $k(\partial h/\partial z +1)|_{z=z_M+\Delta z/2}=R(t)$, where $R(t)$ is the daily rainfall obtained from weather station and is assumed available. 
The lower boundary condition is given by $k(\partial h/\partial z 
+1)|_{z=z_1-\Delta z/2}=k(z_1,t)$.    
According to the existing literature, 
the maximum rooting depth $L_m$ of prairie plants is typically 0.75 to 1.5 meters  and sometimes may reach 3-4 meters~\citep{Coupland1965}, and the value of $\beta$ typically ranges from 0 to 4~\citep{Dong2010}. 
We consider two  sets of true parameter values: $(\beta, L_m) = (1.9, 1.4)$ and $(\beta, L_m) = (1.5, 3.2)$.  
We assume the data is recorded daily at  6 depths (ranging from 0.05 to 0.30 with an increment of 0.05) so that the observed dataset contains 540 data points in total, which clearly exhibits a high degree of nonlinearity. 
The numerical solutions of $t \mapsto f(z, t)$ with $L_m = 1.4$ and $\beta = 1.9$ at six depths  are shown in Figure~\ref{fig:f_sim}. 
To explore the effect of noise on the parameter estimation, we consider multiple choices of the variance of independent Gaussian noise added to the numerical solution of $f$.  
In order to investigate the model's performance with limited data, we repeat the same experiment by reducing the number of observations. For all covariates other than $(\beta, L_m)$, we adopt the  values used in \cite{Dong2010} and keep them fixed throughout the simulation. 

\begin{figure}
\centering
\includegraphics[width=14cm]{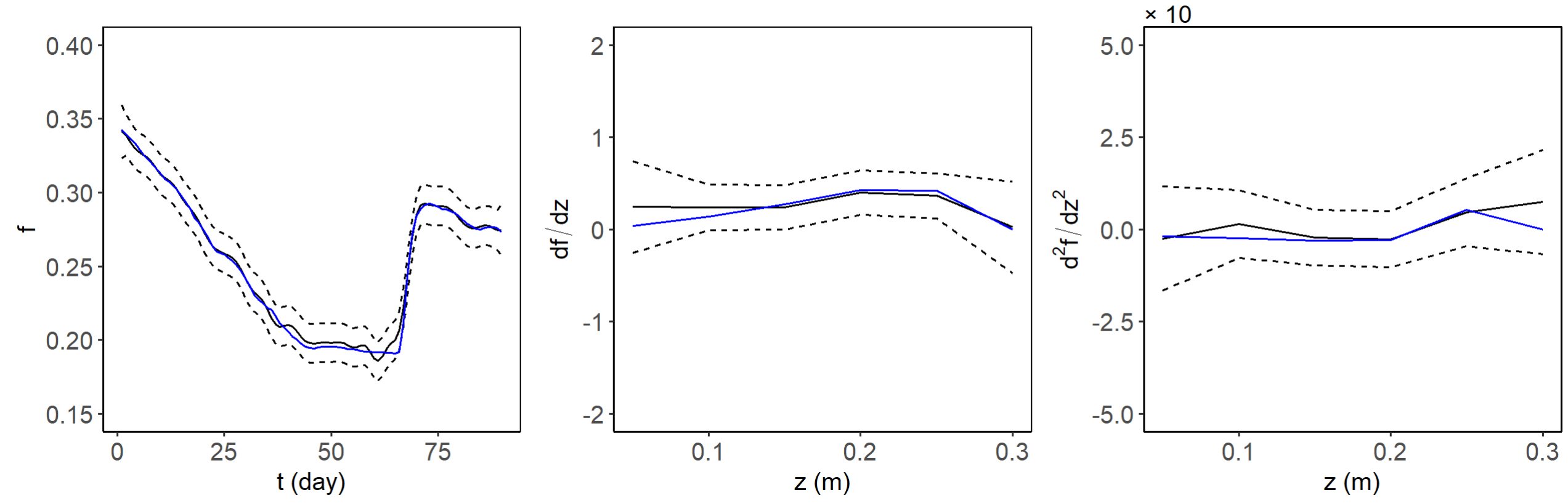}
\caption{A numerical example of the estimates of $f$ and its derivatives obtained by fitting a Gaussian process regression model to $6 \times 90 = 540$ noisy observations, where the error variance is set to 0.05 times the sample   variance of the state variables. 
Left: soil water content $f(z, t)$ at depth $z = 0.3$. 
Middle: $\partial f/\partial z$ at $t= 85 \times 86400$ (i.e., the 85th day).  
Right: $\partial^2 f/\partial z^2$ at $t= 85 \times 86400$. 
Blue solid lines are the true values. Black solid lines are the estimates and dashed lines represent 95\% predictive intervals.}  
 \label{fig:est_f}
\end{figure}

\subsection{Implementation of the methods}
We now describe the implementation details of  our proposed GPC methods. When fitting the Gaussian process to the observed data, we use the Gaussian kernel with kernel hyperparameters estimated by maximum likelihood via the L-BFGS-B optimization algorithm. 
The explicit expression for the covariance function of this predictive distribution is given in Appendix~\ref{Appen:covfun}. A numerical example for the predictive mean estimates of $f$ and its derivatives is  visualized in Figure~\ref{fig:est_f}. 
While overall the Gaussian process estimates are very close to the truth, it can  be seen from the plots that as $z$ gets closer to the boundaries, the estimates  of $f$ and its derivatives become less stable with relatively high standard errors.   
The approximate posterior distribution $\pi_{\rm{GP}}^N$ is defined by \eqref{eq:gpposteriorave} with $N = 100$ and $n_s = 10$, where the $n_s = 10$ design points are selected randomly and $\bm{\omega}_{(1)}, \dots, \bm{\omega}_{(n_s)}$ are generated from  the Gaussian process predictive distribution.
We run the Metropolis--Hastings algorithm  for 3000 iterations and collect $n_t = 15$ posterior samples from the last $1500$ iterations after thinning. 
When calculating the true posterior \eqref{eq:exactpost}, the hyperparameters of the inverse-Gamma prior on $\sigma^2$ are set to $\alpha=\eta=1$. For the BO-GPC method,  
we set the hyperparameters in equation~\eqref{eq:pseudo-post} to $\delta=0.05$ and $\tau=3$. For comparison, we consider the simple Bayesian optimization (BO) method with acquisition function~\eqref{eq:EI}.  
Both BO-GPC and BO  are run for 10 iterations  with 5 randomly generated samples of $\bm{\theta}$ used to initialize the Gaussian process model.   
We implement these two Bayesian optimization algorithms in \texttt{R} by employing the \texttt{DiceKriging} package to fit the Gaussian process. For simple BO, the optimization of the EI criterion is performed via the \texttt{DiceOptim} package; see~\cite{Roustant2012} for further details about the two packages.   
Since for all the three methods  we  evaluate the true posterior for 15 times, their computational times are comparable. 

For comparison, we also implement the GPPDE method proposed in \cite{Rai2019}. They simply use the mean estimates of $f$ and its partial derivatives \eqref{eq:condmean} from the fitted Gaussian process model to evaluate the residual $\xi_i = G( (z_i, t_i), \bm{\omega}_i; \btheta)$  at each data point, and they estimate $\bm{\theta}$ by minimizing the sum of squares of residuals,  $\mathrm{SSRE} \coloneqq \bm{\epsilon}^T\bm{\epsilon}$.  
The optimization algorithm they use is the  Nelder--Mead method, which is unconstrained and does not use gradient information. However, since in our case the gradient of  $\mathrm{SSRE}$ with respect to $(\beta, L_m)$ can be evaluated and bounds for both parameters can be estimated using the domain knowledge, we use the L-BFGS-B method for optimization with constraints $L_m \in (1,4)$ and $\beta\in(0.75,3)$. 

\begin{table}[ht]
\centering
\renewcommand{\arraystretch}{1}
\setlength{\tabcolsep}{4pt}  
\begin{tabular}{ll*{6}{c}}  
\toprule
\multicolumn{2}{c}{Rel. error var  $b$}   & \multicolumn{2}{c}{0.02} & \multicolumn{2}{c}{0.05} & \multicolumn{2}{c}{0.10} \\
 \multicolumn{2}{c}{Truth}   &$L_m = 1.4$ &$\beta = 1.9$ &$L_m = 1.4$ &$\beta = 1.9$ &$L_m = 1.4$ &$\beta = 1.9$ \\
\midrule
  \textbf{Mean} & GPPDE &4.00 &0.75 &4.00 &0.75 &3.95 &0.79 \\
& GPC-I   &1.56 &\textbf{2.01} &1.58 &\textbf{2.08} &1.60 &2.08 \\
& BO-GPC  &\textbf{1.50} &2.12 &\textbf{1.51} &2.14 &\textbf{1.46} &\textbf{2.05} \\
& BO    &1.64 &2.49 &1.62 &2.43 &1.67 &2.55 \\
 \textbf{RMSE} & GPPDE &2.60 &1.15 &2.60 &1.15 &2.56 &1.12 \\
 & GPC-I  &0.39 &\textbf{0.62} &0.41 &\textbf{0.59} &0.45 &\textbf{0.58} \\
 & BO-GPC &\textbf{0.32} &\textbf{0.62} &\textbf{0.33} &0.70 &\textbf{0.31} &0.68 \\
& BO    &0.33 &0.80 &0.38 &0.79 &0.35 &0.82 \\
\midrule
 \multicolumn{2}{c}{Truth}  &$L_m = 3.2$ &$\beta = 1.5$ &$L_m = 3.2$ &$\beta = 1.5$ &$L_m = 3.2$ &$\beta = 1.5$  \\
\midrule
 \textbf{Mean} & GPPDE &1.51 &2.47 &1.76 &2.23 &1.69 &2.24 \\
& GPC-I &3.24 &\textbf{1.53} &3.22 &1.54 &\textbf{3.22} &\textbf{1.53} \\
& BO-GPC  &\textbf{3.20} &\textbf{1.53} &\textbf{3.21} &\textbf{1.52} &\textbf{3.17} &1.47 \\
& BO    &3.29 &1.56 &3.32 &1.59 &3.33 &1.61 \\
 \textbf{RMSE} & GPPDE &1.70 &0.99 &1.55 &0.86 &1.58 &0.89 \\
 & GPC-I &\textbf{0.31} &\textbf{0.27} &\textbf{0.34} &\textbf{0.29} &\textbf{0.32} &\textbf{0.26} \\
 &BO-GPC &0.54 &0.47 &0.53 &0.47 &0.55 &0.47 \\
 & BO    &0.52 &0.42 &0.57 &0.47 &0.52 &0.45 \\
\bottomrule
\end{tabular} 
\caption{Parameter estimates of Richards equation using GPPDE, GPC-I, BO-GPC and BO over 100 replicates with different  levels of noise. The error variance is set to $b$ times the sample variance of the simulated state variables. A total of $n=540$ observations, collected at 6 depths over 90 days, are used. }
\label{table:res}
\end{table}

\subsection{Simulation results}  
The results for two sets of true parameter values over $n_{\mathrm{rep}} = 100$ replicates are summarized in  Table~\ref{table:res}. Estimates under sparse observations are displayed in the supplementary materials. The root mean squared error (RMSE) for each parameter $\theta$  is defined as  
$$
\text{RMSE}(\theta)=\left[\frac{1}{n_{\mathrm{rep}}}\sum_i^{n_{\mathrm{rep}}}(\hat{\theta}_i - \theta^*)^2\right]^{1/2},
$$
where   $\theta^*$ denotes the true parameter value, and $\hat{\theta}_i$ denotes the estimate in the $i$-th replicate.   

\begin{figure}[t]
\centering
\includegraphics[width=14cm]{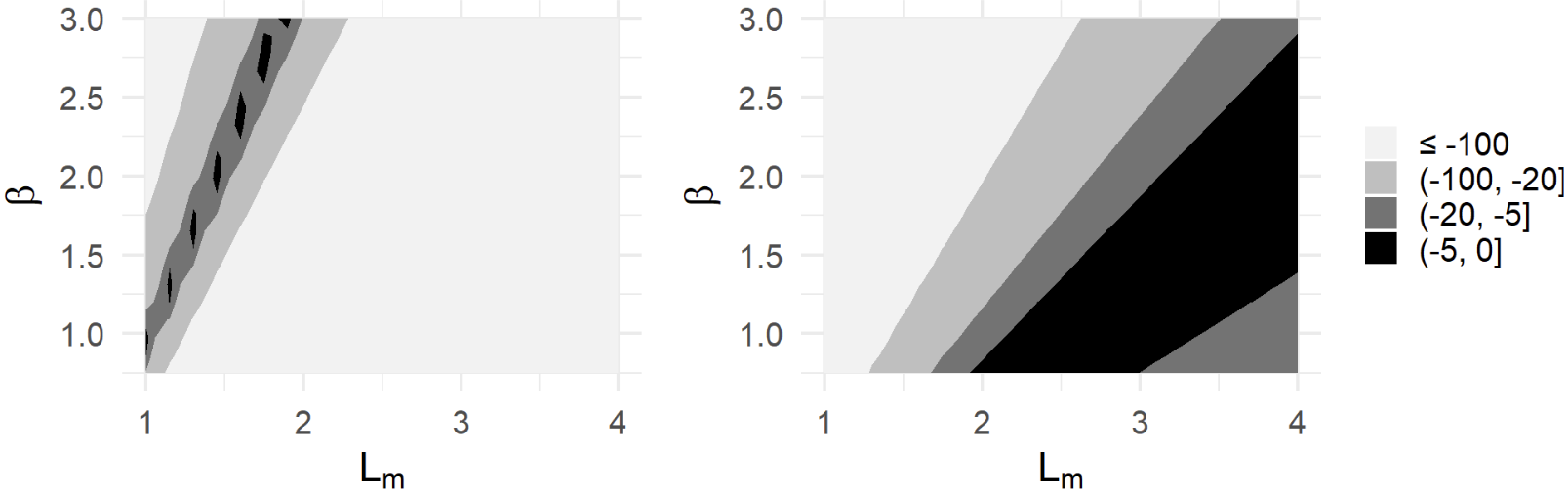}
\caption{Contour plot of the true log-posterior (with maximum value set to zero). 
Left: $L_m=1.4$, $\beta=1.9$; right: $L_m=3.2$, $\beta=1.5$.}
 \label{fig:contour}
\end{figure}

When $L_m=1.4$ and $\beta=1.9$, we find that GPPDE has poor performance compared to the other three methods. This is mainly because the mean estimates of $f$ and its derivatives are inaccurate and volatile in the regions where $f$ is highly nonlinear. Since GPPDE defines the objective function as the unweighted sum of squares of residuals, its performance can be overly sensitive to such estimates. 
In addition, the performance of the optimization algorithm used in GPPDE may largely depend on the initialization since the objective function tends to be multimodal.  
The estimates of BO-GPC have smaller bias than simple BO because of closer mean estimates to the true values. BO-GPC has smaller RMSEs for both parameters than simple BO. These results   reveal that the approximate posterior we define provides useful information on likely values of the parameters. 
The mean estimates of GPC-I are better than BO and comparable to BO-GPC. GPC-I exhibits a larger RMSE for $L_m$, likely due to the irregular shape of the true posterior, which we visualize in Figure~\ref{fig:contour}.  It can be seen that multiple modes lie along a ridge, each attaining a comparably high posterior value. In a few replicates, the proposal distribution given by~\eqref{eq:gpposteriorave} is poorly matched to the true posterior. Consequently, all draws generated from the proposal are away from the ridge, and thus the resulting IS estimate also has a large bias. 
However,  GPC-I has a unique advantage over optimization-based methods in that it may identify multiple local modes other than the true value in one single run, which can offer valuable insights  in practice.  

\begin{figure}[t]
\centering
\includegraphics[width=14cm]{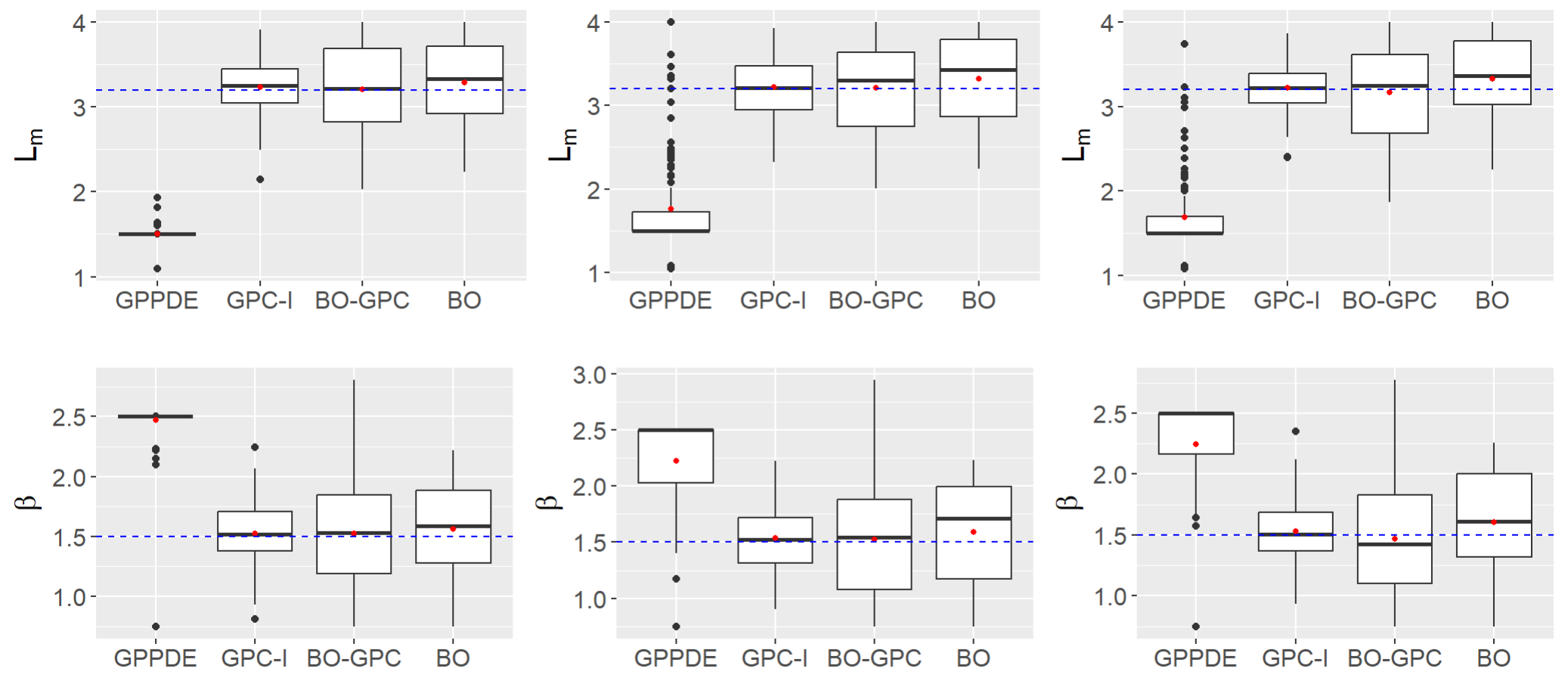}
\caption{Boxplots of the estimates obtained from GPPDE, GPC-I, BO-GPC and BO over 100 replicates at different error levels with $L_m = 3.2$ and $\beta = 1.5$. 
Columns from left to right correspond to relative error variance $b=  0.02, 0.05, 0.1$. The red dots represent the mean estimates of each method, and the true parameter values are indicated by the blue dashed line.  
}
 \label{fig:boxplots}
\end{figure}

When $L_m=3.2$ and $\beta=1.5$, GPPDE still performs significantly worse than the other three methods.  
Further, as shown in the boxplots given in Figure~\ref{fig:boxplots},  the $L_m$ estimate from GPPDE becomes highly unstable as the error variance increases.  
Both Table~\ref{table:res} and Figure~\ref{fig:boxplots}  indicate that  GPC-I clearly achieves the best performance among all four methods,  unlike in the first simulation scenario where BO-GPC tends to outperform GPC-I, especially  in the estimation of $L_m$ 
(in both scenarios, GPC-I yields the most accurate estimate of $\beta$.)  
One possible reason, as suggested by Figure~\ref{fig:contour}, is that the true posterior in this setting is flatter than in the first. Hence, it is easier for the approximate posterior $\pi_{\mathrm{GP}}^N$ to match the overall shape of the true posterior so that $\pi_{\mathrm{GP}}^N$ can act as an efficient proposal distribution providing low-variance IS estimates.   
BO-GPC and BO are comparable, and both perform significantly better than GPPDE. 
The BO-GPC exhibits a slightly larger RMSE than BO. An examination of their search paths suggests that this difference arises from the limited number of iterations used (10 for both methods). Although the approximate posterior is similar in shape to the true posterior, it tends to assign larger posterior mass to suboptimal regions, causing BO-GPC to require slightly more iterations than simple BO to reach the optimum.  

We note that the performance of the proposed methods GPC-I and BO-GPC remain robust across different settings. Regardless of the true parameter values, the error variance does not have a significant impact on their performance, likely because the true posterior is calculated by integrating the error variance out. Moreover, the estimates of GPC-I and BO-GPC also remain relatively stable across different sample sizes; see Appendix~\ref{append:tables} for additional simulation results.

\section{Real field data analysis}
\label{sec:realdata}
\begin{figure}
\centering
\includegraphics[width=14cm]{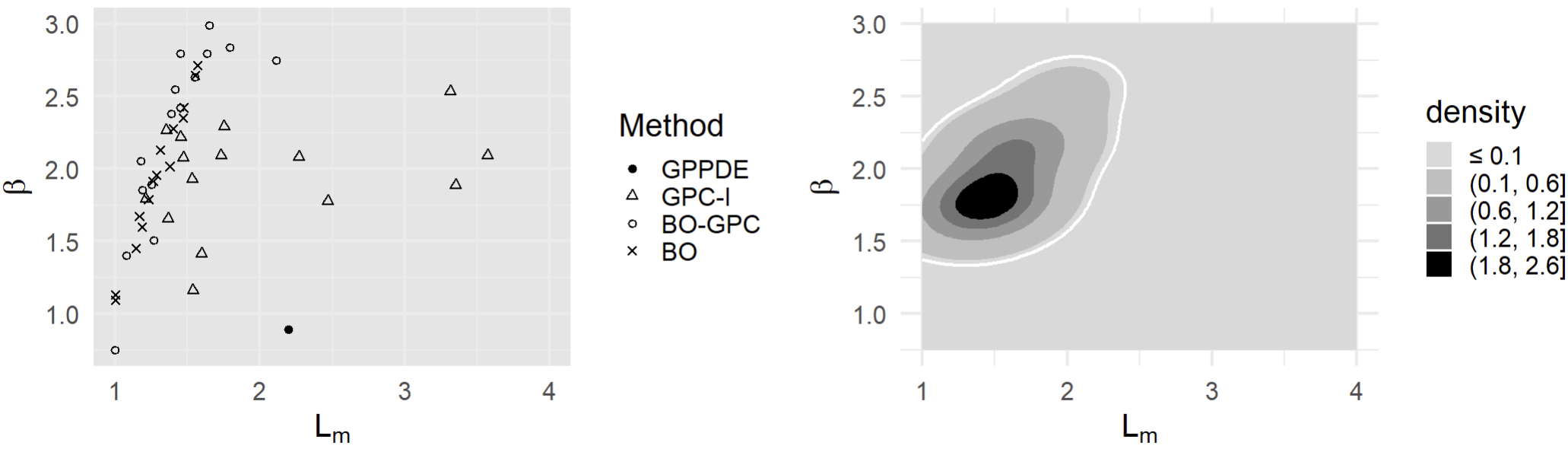}
\caption{Left: Estimates from method GPPDE, GPC-I, BO-GPC, BO. Each method is conducted 15 times. Right: The approximate true posterior density. White contour line contains the 95\% HPD region.} 
 \label{fig:real}
\end{figure}

In this section, we apply the proposed methodology to a real field  data set on a rangeland under moderate grazing by beef cattle~\citep{Dong2010} and estimate the parameters $L_m$ and $\beta$. Soil water contents were measured once a week over 111 days at nine depths (m): 0.067, 0.229, 0.381, 0.533, 0.686, 0.838, 1.067, 1.372, and 1.676. 
For most design points, we recorded two measurements, and the  data set $(\bX, \tilde{\bm{y}})$ was constructed by stacking all observations. As a result, the sample size is $n=260$ in total. The repeated measurements at the same  $(z, t)$ facilitate the estimation for the variance parameter $\sigma^2_y$ in the Gaussian process model described in Section~\ref{sec:GP}. 
The soil parameters in the van Genuchten model are assumed heterogeneous across depths and   determined by the methods described in \cite{Dong2010}; see Appendix~\ref{append:soilpara}. 
When implementing the finite difference scheme for computing the numerical solution to the Richards equation,  time and depth are discretized
with $\Delta t = 400$ and $\Delta z = 0.025$ respectively. 
To determine the initial condition, polynomial functions were fitted to the measured soil water content on the first day. Then the soil water contents at each depth were interpolated and used as the initial condition. The lower boundary condition is identical to that used in the simulation study presented in Section \ref{sec:sim}. The upper boundary condition is set to $k(\partial h/\partial z +1)|_{z=z_{\rm{top}}+\Delta z/2}=R(t)-E_a(t)$ where $z_{\rm{top}}$ is the negative of smallest depth grid point and $E_a(t)$ is the actual soil evaporation computed by formula (14) in \cite{Dong2010}. 
Implementation of the proposed methods are the same as described in \ref{sec:sim},  except that we tune  the temperature parameter for the approximate posterior \eqref{eq:gpposteriorave} to further flatten the posterior, facilitating the exploration of parameter space.  
All the above-mentioned approaches were conducted 15 times and the estimates for the parameters are displayed in the left panel of Figure~\ref{fig:real}.  
In addition, we approximate the true posterior using~\eqref{eq:kde} in one replicate and compute the HPD region which is visualized in the right panel  of Figure~\ref{fig:real}. 
Notably, the HPD region also contains the estimate reported in~\cite{Dong2010}, which was obtained using a conventional grid search method. 
The GPPDE method clearly fails to provide accurate estimates, as its estimates are deterministic and significantly deviate from the high-density region of the true posterior.
GPC-I produces consistent estimates in most runs, but in some runs it exhibits substantial bias due to a mismatch between the approximate and true posterior distributions. 
This behavior aligns with the phenomenon observed in our simulation study. 
BO-GPC and BO yield comparable results and the estimates lie within a region which largely overlaps with the HPD region of GPC-I.

As mentioned in Section~\ref{sec:intro}, existing methods for inference with the Richards equation typically take significant amount of computing time, and may   run the risk of ``incomplete tuning"  due to the existence of several local minima~\citep{Janssen1995,Rasiah1992}. Our results indicate that the proposed GPC methods  significantly improve the efficiency and robustness in terms of estimating $L_m$ and $\beta$. Further, GPC-I can quantify the uncertainty of our estimates by constructing the credible region. 
The significance of these improvements is particularly notable in agricultural sciences, since both $L_m$ and $\beta$ are known to be difficult to estimate   under field conditions. Even with substantial investment of researchers' time and labor, the resulting estimates are typically based on small sample sizes and subject to large uncertainty~\citep{Coupland1965,chandra1996}.  
The simulation study and real data analysis demonstrates the potential usefulness of the proposed methods in situations where available observations are limited, which is common in agricultural studies on many soil-plant systems. 
Furthermore, while this work focuses on estimating $L_m$ and $\beta$, our methods are also applicable to inferring other soil parameters, including those treated as known in this study. 

\section{Concluding remarks}
\label{sec:conclusion}
In this paper, we propose a new Gaussian process collocation method for estimating unknown parameters in PDE models from observed noisy data, together with a novel correction procedure that leverages numerical solvers when they are available. 
Through statistical inference with the Richards equation using both simulated and real data sets, our method demonstrates both computational efficiency and robustness across various settings, outperforming existing approaches. The proposed framework is broadly applicable to general differential equations that arise across different scientific domains. 

Several directions can be pursued to extend this work. First, when the gradient information of the approximate posterior is available, Metropolis-Adjusted Langevin Algorithm~\citep{Gareth1998} can be employed to improve the sampling efficiency. In addition, algorithms like simulated tempering~\citep{Marinari1992} can be adopted to enhance mode exploration of highly multimodal posterior distributions. Second, the proposed method can be extended to estimate unknown parameters in systems of PDEs (e.g., the Lotka–Volterra equations). 
The main modification to the current procedure is to model the state variables jointly via multiple Gaussian processes and introduce additional hyperparameters to characterize their correlation. An  example of this approach is provided in~\cite{Bonilla2007}, which may be integrated into the proposed GPC methods. 
\section*{Acknowledgments} 
The authors are grateful to Jun Yang at the University of Copenhagen for inspiring this research and helpful discussions.
\section*{Funding}
This work was supported in part by the NSF grants DMS-2311307 and DMS-2245591, the Cotton Incorporated grant 20-557TX, and the USDA-NIFA Hatch Project grant 9574–2.

\bibliographystyle{apalike} 
\bibliography{draft}   
\newpage
\appendix
\section{Appendix}
\subsection{Derivatives of the log marginal likelihood}
\label{Appen:derimag}
The log marginal likelihood of the fitted GP is given by
\begin{align}
\label{eq:logmargin}
    \log p(\mathbf{\tilde{y}}\,|\,\mathbf{X},\bm{\phi})=&-\frac{1}{2}(\mathbf{\tilde{y}}-m_y(\mathbf{X}))^T\mathbf{K}_y(\mathbf{\tilde{y}}-m_y(\mathbf{X}))\\
    &-\frac{1}{2}\log|\mathbf{K}_y|-\frac{n}{2}\log2\pi,
\end{align}
where $\bm{\phi}=\{l_1,\dots,l_m,\sigma_s^2,\sigma^2_y\}$. Let $\bm{\zeta}=\log\bm{\phi}$ be the logarithms of the hyperparameters, then the partial derivatives of \eqref{eq:logmargin} with respect to each $\bm{\zeta}_i$ is
\begin{align*}
    \frac{\partial}{\partial \bm{\zeta}_h}\log p(\mathbf{\tilde{y}}\,|\,\mathbf{X},\bm{\phi})=&\frac{1}{2}(\mathbf{\tilde{y}}-m_y(\mathbf{X}))^T\mathbf{K}^{-1}_y\frac{\partial\mathbf{K}_y}{\partial\bm{\zeta}_h}\mathbf{K}_y^{-1}(\mathbf{\tilde{y}}-m_y(\mathbf{X}))\\
    -&\frac{1}{2}\text{tr}(\mathbf{K}_y^{-1}\frac{\partial\mathbf{K}_y}{\partial\bm{\zeta}_h});\quad h=1,\dots,m+2,
\end{align*}
where $(\mathbf{K}_y)_{ij}=\sigma_s^2\exp(-\frac{1}{2}\sum^m_{d=1}\frac{(x_{di}-x_{dj})^2}{l_d^2})+\delta_{ij}\sigma_y^2$
and $\delta_{ij}$ is the Kronecker delta. Then, we have
\begin{align*}
    \left(\frac{\partial\mathbf{K}_y}{\partial\bm{\zeta}_h}\right)_{ij}&=\sigma_s^2\exp(-\frac{1}{2}\sum^m_{d=1}\frac{(x_{di}-x_{dj})^2}{l_d^2})\frac{(x_{hi}-x_{hj})^2}{l_h^2};\quad 1\leq h\leq m,\\
\left(\frac{\partial\mathbf{K}_y}{\partial\bm{\zeta}_{m+1}}\right)_{ij}&= \sigma_s^2\exp(-\frac{1}{2}\sum^m_{d=1}\frac{(x_{di}-x_{dj})^2}{l_d^2}),\\
\left(\frac{\partial\mathbf{K}_y}{\partial\bm{\zeta}_{m+2}}\right)_{ij}&= \delta_{ij}\sigma_y^2.
\end{align*}

\subsection{Derivatives of the Richards equation}
\label{Appen:ricfact}
The Richards equation can be written as:
\begin{align}
    \frac{\partial f}{\partial t}&=\frac{\partial}{\partial z}\left[k\frac{\partial h}{\partial z}\right]+\frac{\partial k}{\partial z}-S(h,z,t;\bm{\theta})\\
    &=\frac{\partial k}{\partial z}\frac{\partial h}{\partial z}+k\frac{\partial^2 h}{\partial z^2}+\frac{\partial k}{\partial z}- S(h,z,t;\bm{\theta}).
\end{align}
In this appendix, we show how these four parts on the right hand side of the equation can be expressed in terms of $f$ and its partial derivatives.
According to \eqref{eq:vanGa} of van Genuchten model, we write $h$ as a function of $f$:
\begin{equation}
\label{eq:f-to-h}
    h=-\frac{1}{\gamma}\left[\left(\frac{c_s-c_r}{f-c_r}\right)^{1/\nu}-1\right]^{1-\nu}.
\end{equation}
Thus, the sink term $S$ can be written as a function of $f$, $z$, $t$ and $\bm{\theta}$.\newline
\\
For $\partial k/\partial z$, we first obtain $\partial k/\partial f$ from \eqref{eq:vanGb} and \eqref{eq:vanGc}. Then we express it as the following form by the chain rule:
\begin{align}
\label{eq:dkdz}
    \frac{\partial k}{\partial z} =&\frac{\partial k}{\partial f}\frac{\partial f}{\partial z}\notag\\
    =&\frac{k_{\rm{sat}}}{c_s-c_r}
    \Bigg\{2C^{1/2}\left[1-(1-C^{1/\nu})^{\nu}\right]\left[(1-C^{1/\nu})^{\nu-1}(C^{1/\nu-1})\right]\notag\\
    &\qquad\qquad +\frac{1}{2}C^{-1/2}\left[1-(1-C^{1/\nu})^{\nu}\right]^2\Bigg\}\frac{\partial f}{\partial z}.
\end{align}
For $(\partial k/\partial z)\partial h/\partial z$, we first calculate $\partial h/\partial f$ from \eqref{eq:f-to-h} and have
\begin{align}
\label{eq:dkhdz2}
    \frac{\partial k}{\partial z}\frac{\partial h}{\partial z}&=\frac{\partial k}{\partial z}\frac{\partial h}{\partial f}\frac{\partial f}{\partial z}\notag\\
    &=\frac{\partial k}{\partial z}\frac{1-\nu}{\gamma \nu(c_s-c_r)}C^{-(1+\nu)/\nu}\left\{C^{-1/\nu}-1\right\}^{-\nu}\frac{\partial f}{\partial z}.
\end{align}
Replace $\partial k/\partial z$ by equation \eqref{eq:dkdz} and obtain the desired form. \newline
\\
For $k(\partial^2 h/\partial z^2)$, $k$ can be substituted by \eqref{eq:vanGb} and we deal with $\partial^2 h/\partial z^2$ as follows:
\begin{align}
    \frac{\partial^2 h}{\partial z^2}=\frac{\partial}{\partial z}\left(\frac{\partial h}{\partial z}\right)=\frac{\partial}{\partial z}\left(\frac{\partial h}{\partial f}\frac{\partial f}{\partial z}\right)=\frac{\partial^2 h}{\partial f\partial z}\frac{\partial f}{\partial z}+\frac{\partial h}{\partial f}\frac{\partial^2 f}{\partial z^2}.
\end{align}
$\partial h/\partial f$ has been calculated in \eqref{eq:dkhdz2} and 
\begin{align}
   \frac{\partial^2 h}{\partial f\partial z}&= \frac{\partial}{\partial C}\left(\frac{\partial h}{\partial f}\right)\frac{\partial C}{\partial f}\frac{\partial f}{\partial z}\\
   &=\frac{1-\nu}{\gamma \nu(c_s-c_r)^2}\frac{\partial f}{\partial z}\Bigg\{-\left(\frac{1}{\nu}+1\right)C^{-(1+2\nu)/\nu}\left(C^{-1/\nu}-1\right)^{-\nu}\\
   &\qquad\qquad\qquad\qquad\qquad +C^{-2(\nu+1)/\nu}\left(C^{-1/\nu}-1\right)^{-\nu-1}\Bigg\}.
\end{align}

\subsection{Covariance functions in predictive distributions for the Richards equation}
\label{Appen:covfun}
 Gaussian covariance function is given by 
\begin{equation}
k(\mathbf{x}_i,\mathbf{x}_j)=\sigma_s^2\exp\left\{-\frac{1}{2}\sum^m_{l=1}\frac{(x_{i,l}-x_{j,l})^2}{l_l^2}\right\}.
\end{equation}
In Richards equation, first and second partial derivatives of $z$, first partial derivatives of $t$ are included. Assuming that $x_d$ represents  $z$ and $x_e$ represents  $t$, for $i,j=1,2,...,n$, we have
$$\frac{\partial k(\mathbf{x}_i,\mathbf{x}_j)}{\partial x_{j,d}}=k(\mathbf{x}_i,\mathbf{x}_j)\frac{x_{i,d}-x_{j,d}}{l_d^2},$$
$$\frac{\partial^2 k(\mathbf{x}_i,\mathbf{x}_j)}{\partial x_{i,e}x_{j,d}}=\frac{k(\mathbf{x}_i,\mathbf{x}_j)}{l_e^2}\left[\delta_{ed}-\frac{(x_{i,d}-x_{j,d})(x_{i,e}-x_{j,e})}{l_d^2}\right],$$
$$
\frac{\partial^2 k(\mathbf{x}_i,\mathbf{x}_j)}{\partial x_{j,d}^2}=\frac{k(\mathbf{x}_i,\mathbf{x}_j)}{l_d^2}\left[\frac{(x_{i,d}-x_{j,d})^2}{l_d^2}-1\right],
$$
$$
\frac{\partial^3 k(\mathbf{x}_i,\mathbf{x}_j)}{\partial x_{i,e}x_{j,d}^2}=\frac{k(\mathbf{x}_i,\mathbf{x}_j)(x_{i,e}-x_{j,e})}{l_e^2l_d^2}\left[2\delta_{ed}+1-\frac{(x_{i,d}-x_{j,d})^2}{l_d^2}\right],
$$
and
$$
\frac{\partial^4 k(\mathbf{x}_i,\mathbf{x}_j)}{\partial x_{i,d}^2x_{j,d}^2} =\frac{k(\mathbf{x}_i,\mathbf{x}_j)}{l_d^4}\left[3-\frac{6(x_{i,d}-x_{j,d})^2}{l_d^2}+\frac{(x_{i,d}-x_{j,d})^4}{l_d^4}\right].
$$

\subsection{Details of the sink term modeling}
\label{Appen:sink}
In this study, we model the sink term by
\begin{equation}
     S(h, z, t;\bm{\theta}) = \left\{\begin{array}{cc}
      0,    & \quad \text{ if }  z > L_m Z_r(t), \\
      \frac{(1 + \beta) \psi(z, t)  }{L_m Z_r(t)}\left( 1 - \frac{|z|}{L_m Z_r(t)}\right)^\beta,   &  \quad \text{ if }  z \leq L_m Z_r(t), 
      \end{array}  
     \right.  
\end{equation}
 where $L_m Z_r(t)=L_r(t)$ is the actual plant rooting depth (m) at time $t$,
 $$\psi(z, t)=\alpha(h, \bm{A}(t))  T_p(t).$$
 $T_p(t)$ is the potential transpiration ($\text{m}\cdot \text{s}^{-1}$) at time $t$ calculated according to \cite{Ritchie1972}, for agricultural crops with leaf area index ($\text{m}^2\cdot \text{m}^{-2}$) less than 2.7. 
And $\alpha$ is known as the Feddes reduction function~\citep{Feddes1978}, which is a piecewise linear function ranging from 0 and 1 with parameter $\bA(t) = (A_1(t), A_2(t), A_3(t), A_4(t) )$. It has the following form: 
 \begin{equation}\label{eq:alpha}
    \alpha(h, \bA) = \left\{\begin{array}{cc}
      0,   &  \text{ if } h < A_4 \text{ or } h \geq A_1,  \\
      \frac{h - A_4}{A_3 - A_4},   &  \text{ if } A_4 \leq h < A_3,  \\
      1,   &  \text{ if } A_3 \leq h < A_2,  \\
      \frac{A_1 - h}{A_1 - A_2},   &  \text{ if } A_2 \leq h < A_1.       
    \end{array}
    \right.
\end{equation} 
  where $A_1$ and $A_4$ are the soil matric potentials at anaerobiosis point and at permanent wilting point (m), respectively; $A_2$ and $A_3$ mark the range of water potential for optimal root extraction. $L_m$ is the maximum rooting depth (m) and $Z_r(t)$ is the sigmoidal function of \citet{Gardner1985} used to describe the root growth. To calculate $Z_r(t)$, we specify the start day of root growth, the days required for the maturity of plants and some empirical coefficients determining the shape of the growth curve.
Finally, $$(1 + \beta)\left( 1 - \frac{|z|}{L_m Z_r(t)}\right)^\beta$$ is called the normalized root density distribution.
Following the approach discussed in \cite{Dong2010}, $Z_r(t)$, $\alpha(h,\bm{A})$ and $T_p(t)$ can be determined. 

\subsection{A Picard's iteration method for solving the Richards equation}
\label{Appen:picard}
Write the Richards equation as 
\begin{equation} 
\label{eq:hrich}
    \frac{\partial f}{\partial t} = \frac{\partial}{\partial z} \left[k \frac{\partial h}{\partial z} \right] + \frac{\partial k}{\partial z} - S,
\end{equation}
where $h$ is a function of $z$, $t$ and $S$ is a function $z, t, h$. We view $f, k$ as functions of $h$ given the van Genuchten model. Fix some time point $t$ and the time step size $\Delta t$. For simplicity, write  
\begin{align*}
    \tilde{h} (z) = h(z, t),   \quad h(z) = h(z, t + \Delta t), 
\end{align*}
and let  $\tilde{f}(z) = f(\tilde{h}(z))$, $f(z) = f(h(z))$, and $k(z) = k(h(z))$. Discretizing~\eqref{eq:hrich} backwards in time we get \begin{equation}\label{eq:richards.time.diff}
    \frac{f - \tilde{f} }{ \Delta t} 
    = \frac{\partial}{\partial z}\left[ k \frac{ \partial h }{ \partial z} \right]
    + \frac{\partial k} {\partial z} -  S. 
\end{equation} 
By solving~\eqref{eq:richards.time.diff}, we get $h$ at time $t + \Delta t$, and repeating this calculation we get $h$ over the time domain. $f$ can be then calculated according to~\eqref{eq:vanGa}. When $t=0$, $\tilde{h}$ is determined by the initial condition.

To solve~\eqref{eq:richards.time.diff}, we use a mixed form Picards method, which creates a sequence of solutions $h^0, h^1, \dots$ that converges to the solution to~\eqref{eq:richards.time.diff}. 
We can simply set $h^0 = \tilde{h}$. 
Let $\upsilon^j = h^{j} - h^{j-1}$ be the increment at the $j$-th iteration, which should eventually converge to zero. 
Define $c(h) = \partial f / \partial h$, which admits a closed-form expression. Let $c^j = c(h^j)$ and define $f^{j}, k^{j}, S^{j}$ similarly. For sufficiently small $\upsilon^j$, we should have 
\begin{equation}
    f^{j}  \approx f^{j - 1} + c^{j - 1} \upsilon^j. 
\end{equation}
Given $h^{j - 1}$, we calculate $f^{j-1}, k^{j-1}, S^{j-1}, c^{j-1}$ and find $\upsilon^j$ by solving 
\begin{equation}\label{eq1}
    \frac{f^{j - 1} + c^{j - 1} \upsilon^j - \tilde{f} }{ \Delta t} 
    = \frac{\partial}{\partial z}\left( k^{j-1} \left\{\frac{ \partial (h^{j-1} + \upsilon^j) }{ \partial z}  + 1\right\} \right)
     -  S^{j-1}.   
\end{equation}
Discretize $z$ by considering points $z_1, z_2, \dots, z_M$. We define $h_i^j = h^j(z_i)$ and $c^j_i$, $k_i^j$, $f^j_i$, $S^j_i$ in the same way. Let $\upsilon^j_i$ be the increment of $h$ at $j$-th iteration at $z_i$. First, we focus on the interior points (i.e., points other than $z_1, z_M$). Standard approximation to first and second derivatives yields that, for $i = 2, \dots, M-1$, 
\begin{align*}
     & \frac{\partial}{\partial z}\left( k^{j-1} \left\{\frac{ \partial (h^{j-1} + \upsilon^j) }{ \partial z}  + 1\right\} \right)\Big|_{z = z_i} \\
     \approx\; & \frac{1}{\Delta z} \Bigg(  k^{j-1}_{i+0.5}\left\{  \frac{ h^{j-1}_{i + 1} + \upsilon^j_{i+ 1} 
    - h^{j-1}_{i } - \upsilon^j_{i } }{\Delta z} + 1 \right\}  
    - \\
    & k^{j-1}_{i-0.5}\left\{  \frac{ h^{j-1}_{i} + \upsilon^m_{i} 
    - h^{j-1}_{i-1} - \upsilon^j_{i-1} }{\Delta z} + 1 \right\} \Bigg). 
\end{align*}
Then, we obtain that 
\begin{equation}
\label{eq:picint}
    A_i \upsilon_{i-1}^j + B_i \upsilon_i^j + C_i \upsilon_{i+1}^j = D_i, 
\end{equation}
where 
\begin{align*}
    A_i =\;& - \frac{ k_{i-0.5}^{j-1} }{ \Delta z^2}, \quad
    B_i = \frac{ c_{i}^{j - 1}  }{ \Delta t} + \frac{ k_{i+0.5}^{j-1}   + k_{i-0.5}^{j-1}  }{\Delta z^2},\quad 
    C_i = - \frac{ k_{i+0.5}^{j-1} }{ \Delta z^2},  \\
    D_i =\;& \frac{ k_{i+0.5}^{j-1} (h_{i+1}^{j-1} - h_{i}^{j-1}) - k_{i-0.5}^{j-1} (h_{i}^{j-1} - h_{i-1}^{j-1}) }{\Delta z^2}   
      + \frac{ k_{i+0.5}^{j-1} - k_{i-0.5}^{j-1} }{\Delta z} -     \frac{f_{i}^{j-1}- \tilde{f}_{i} }{ \Delta t}\\
      &- S^{j-1}_i.  
\end{align*}
Now, we find the equations at $z_1$ and $z_M$, which will involve two boundary conditions. We write $G_{M+0.5}=k(\partial h/\partial z +1)|_{z=z_M+\Delta z/2}$ and $G_{0.5}=k(\partial h/\partial z 
+1)|_{z=z_1-\Delta z/2}$. Similarly, we have \begin{align*}
     & \frac{\partial}{\partial z}\left( k^{j-1} \left\{\frac{ \partial (h^{j-1} + \upsilon^j) }{ \partial z}  + 1\right\} \right)\Big|_{z = z_1} \\
     \approx\; & \frac{1}{\Delta z} \left(  k^{j-1}_{1.5}\left\{  \frac{ h^{j-1}_{2} + \upsilon^j_{2} 
    - h^{j-1}_{1} - \upsilon^j_{1} }{\Delta z} + 1 \right\}  
    - G_{0.5}^{j-1} \right), 
\end{align*}
and  
\begin{align*}
     & \frac{\partial}{\partial z}\left( k^{j-1} \left\{\frac{ \partial (h^{j-1} + \upsilon^j) }{ \partial z}  + 1\right\} \right)\Big|_{z = z_M} \\
     \approx\; & \frac{1}{\Delta z  } \left( G_{M+0.5}^{j-1} -   k^{j-1}_{M-0.5}\left\{  \frac{ h^{j-1}_{M} + \upsilon^j_{M}     - h^{j-1}_{M-1} - \upsilon^j_{M-1} }{\Delta z} + 1 \right\}  
     \right). 
\end{align*}
So we get 
\begin{align}
\label{eq:picbound}
\begin{aligned}
    B_1 \upsilon_1^j + C_1 \upsilon_2^j =\;& D_1, \\
    A_M \upsilon_{M-1}^j + B_M \upsilon_M^j  =\;& D_M, 
\end{aligned}
\end{align}
where  
\begin{align*}
    B_1 =\,&  \frac{c_{1}^{j-1} }{ \Delta t} + \frac{  k_{1.5}^{j-1} }{ \Delta z^2},  \quad
    C_1 = - \frac{  k_{1.5}^{j-1} }{ \Delta z^2}, \\
    D_1 =\;&  \frac{  k_{1.5}^{j-1} (h_{2}^{j-1}  - h_{1}^{j-1})  }{ \Delta z^2}   +  \frac{   (k_{1.5}^{j-1}  - G_{0.5}^{j-1} ) }{\Delta z} - \frac{f_{1}^{j-1}   - \tilde{f}_{1}}{\Delta t}  - S_1^{j-1}, 
\end{align*}
\begin{align*}
    A_M =\, & - \frac{ k_{M - 0.5}^{j-1} }{ \Delta z^2},\quad
    B_M =  \frac{c_{M}^{j-1} }{ \Delta t} + \frac{ k_{M - 0.5}^{j-1} }{ \Delta z^2}, \\ 
    D_M =\;&  - \frac{  k_{M - 0.5}^{j-1} (h_{M}^{j-1}  - h_{M-1}^{j-1})  }{\Delta z^2}   +  \frac{ (G_{M+0.5}^{j-1} - k_{M - 0.5}^{j-1}   ) }{ \Delta z} - \frac{f_{M}^{j-1}   - \tilde{f}_{M}}{\Delta t} - S_{M}^{j-1}. 
\end{align*}
Then, according to~\eqref{eq:picint} and~\eqref{eq:picbound}, we have a tridiagonal system of equations for $\upsilon^j_i$, which can be solved by the Thomas algorithm.

\subsection{Soil parameters used in real field data analysis}
\label{append:soilpara}
The soil parameters used in real field data analysis are given by
\begin{table}[H]
\centering
\label{table3}
\begin{tabular}{ccccccc}
\toprule
Depth (m) & $c_r$ & $c_s$ & $\gamma$ (m$^{-1}$) & $\nu$ & $k_{\text{sat}}$ ($\text{m}\cdot \text{s}^{-1}$) \\
\midrule
(0, 0.15] & 0.156 & 0.600 & 5.870 & 0.273 & $2.82 \times 10^{-5}$ \\
(0.15, 0.30] & 0.156  & 0.600& 5.870 & 0.273  & 6.73 $\times$ 10$^{-6}$  \\
(0.30, 0.46] & 0.001  & 0.560  & 0.362  &0.171  & 1.46 $\times$ 10$^{-6}$ \\
(0.46, 0.61] & 0.001  & 0.560& 0.362  & 0.171  & 3.50 $\times$ 10$^{-7}$ \\
(0.61, 0.76] & 0.001 & 0.560 & 0.362  &0.171 & 8.36 $\times$ 10$^{-8}$ \\
(0.76, 3.00]& 0.001 & 0.560 & 0.362 & 0.171 & $2.00 \times 10^{-8}$ \\
\bottomrule
\end{tabular}
\caption{Soil parameters used in field measured data analysis}
\end{table}
\newpage
\subsection{Simulation results}
\label{append:tables}
In this section. we present additional simulation results. 

\begin{table}[!h]
\centering
\label{table1}
\renewcommand{\arraystretch}{1.2}
\setlength{\tabcolsep}{4pt} 
\resizebox{\textwidth}{!}{
\begin{tabular}{l ll*{6}{c}}  
\toprule
& Rel. error var $b$&  & \multicolumn{2}{c}{0.02} & \multicolumn{2}{c}{0.05} & \multicolumn{2}{c}{0.10} \\
 $n$ & Truth & &$L_m=1.4$&$\beta=1.9$&$L_m=1.4$&$\beta=1.9$&$L_m=1.4$&$\beta=1.9$\\
\midrule
\multirow{8}{*}{\textbf{540 (6$\times$90)} \quad} & \textbf{mean} & GPPDE &4.00 &0.75 &4.00 &0.75 &3.95 &0.79 \\
& & GPC-I   &1.56 &2.01 &1.58 &2.08 &1.60 &2.08 \\
& & BO-GPC  &1.50 &2.12 &1.51 &2.14 &1.46 &2.05 \\
& & BO    &1.64 &2.49 &1.62 &2.43 &1.67 &2.55 \\

 & \textbf{RMSE} & GPPDE &2.60 &1.15 &2.60 &1.15 &2.56 &1.12 \\
& & GPC-I  &0.39 &0.62 &0.41 &0.59 &0.45 &0.58 \\
& & BO-GPC &0.32 &0.62 &0.33 &0.70 &0.31 &0.68 \\
& & BO    &0.33 &0.80 &0.38 &0.79 &0.35 &0.82 \\
\midrule
\multirow{8}{*}{\textbf{270 (3$\times$90)}} & \textbf{mean} & GPPDE &4.00 &0.75 &4.00 &0.75 &4.00 &0.75 \\
& & GPC-I  &1.56 &2.08 &1.60 &2.16 &1.50 &2.06 \\
& &BO-GPC &1.46 &2.12 &1.48 &2.12 &1.46 &2.11 \\
& & BO    &1.67 &2.56 &1.64 &2.48 &1.62 &2.44 \\

 & \textbf{RMSE} & GPPDE &2.60 &1.15 &2.60 &1.15 & 2.6&1.15 \\
& & GPC-I  &0.48 &0.56 &0.50 &0.62 &0.33 &0.60 \\
& &BO-GPC  &0.33 &0.71 &0.33 &0.74 &0.32 &0.65 \\
& & BO    &0.34 &0.83 &0.35 &0.84 &0.32 &0.79 \\
\midrule
\multirow{8}{*}{\textbf{135 (3$\times$45)}} & \textbf{mean} & GPPDE &3.99 &0.75 &3.95 &0.75 &3.94 &0.75 \\
& & GPC-I  &1.90 &2.15 &1.72 &2.07 &1.74 &2.12 \\
& &BO-GPC &1.48 &2.12 &1.51 &2.22 &1.44 &2.07 \\
& & BO    &1.64 &2.50 &1.62 &2.44 &1.63 &2.46 \\

 & \textbf{RMSE} & GPPDE &2.59 &1.15 &2.55 &1.15 &2.56 &1.15 \\
& & GPC-I  &0.85 &0.58 &0.71 &0.55 &0.68 &0.53 \\
& &BO-GPC &0.30 &0.63 &0.29 &0.68 &0.28 &0.66 \\
& & BO    &0.34 &0.84 &0.32 &0.77 &0.32 &0.77 \\
\bottomrule
\end{tabular}}
\caption{Parameter estimates of Richards equation using GPPDE, GPC-I, BO-GPC and BO over 100 replicates with different number of observations and levels of noise when $L_m=1.4$ and $\beta=1.9$. The error variance is set to $b$ times the sample variance of the simulated state variable. Observations at 6 depths across 90 days are used. Estimates under fewer observations are also included. 
}
\end{table}
\newpage

\begin{table}
\centering
\label{table2}
\renewcommand{\arraystretch}{1.2}
\setlength{\tabcolsep}{4pt} 
\resizebox{\textwidth}{!}{
\begin{tabular}{l ll*{6}{c}}  
\toprule
& Rel. error var $b$&  & \multicolumn{2}{c}{0.02} & \multicolumn{2}{c}{0.05} & \multicolumn{2}{c}{0.10} \\
$n$& Truth & &$L_m=3.2$&$\beta=1.5$&$L_m=3.2$&$\beta=1.5$&$L_m=3.2$&$\beta=1.5$\\
\midrule
\multirow{8}{*}{\textbf{540 (6$\times$90)}} & \textbf{mean} & GPPDE &1.51 &2.47 &1.76 &2.23 &1.69 &2.24 \\
& & GPC-I &3.24 &1.53 &3.22 &1.54 &3.22 &1.53 \\
& & BO-GPC  &3.20 &1.53 &3.21 &1.52 &3.17 &1.47 \\
& & BO    &3.29 &1.56 &3.32 &1.59 &3.33 &1.61 \\

 & \textbf{RMSE} & GPPDE &1.70 &0.99 &1.55 &0.86 &1.58 &0.89 \\
& & GPC-I &0.31 &0.27 &0.34 &0.29 &0.32 &0.26 \\
& &BO-GPC &0.54 &0.47 &0.53 &0.47 &0.55 &0.47 \\
& & BO    &0.52 &0.42 &0.57 &0.47 &0.52 &0.45 \\
\midrule
\multirow{8}{*}{\textbf{270 (3$\times$90)}} & \textbf{mean} & GPPDE &1.50 &2.50 &1.50 &2.50 &1.50 &2.47 \\
& & GPC-I  &3.24 &1.54 &3.28 &1.58 &3.26 &1.57 \\
& &BO-GPC  &3.17 &1.43 &3.25 &1.54 &3.21 &1.55 \\
& & BO    &3.31 &1.59 &3.34 &1.61 &3.23 &1.52 \\

 & \textbf{RMSE} & GPPDE &1.70 &1.00 &1.70 &1.00 &1.70 &1.00 \\
& &  GPC-I  &0.22 &0.22 &0.21 &0.21 &0.26 &0.26 \\
& &BO-GPC  &0.56 &0.47 &0.59 &0.51 &0.58 &0.52 \\
& & BO    &0.52 &0.44 &0.55 &0.46 &0.56 &0.46 \\
\midrule
\multirow{8}{*}{\textbf{135 (3$\times$45)}} & \textbf{mean} & GPPDE &1.70 &0.84 &2.20 &1.04 &2.77 &1.13 \\
& &  GPC-I  &3.21 &1.57 &3.21 &1.58 &3.21 &1.61 \\
& &BO-GPC &3.14 &1.35 &3.31 &1.58 &3.19 &1.49 \\
& & BO    &3.28 &1.56 &3.40 &1.67 &3.22 &1.51 \\

 & \textbf{RMSE} & GPPDE &1.59 &0.72 &1.44 &0.78 &1.23 &0.80 \\
& & GPC-I  &0.23 &0.26 &0.19 &0.23 &0.16 &0.22 \\
& &BO-GPC &0.60 &0.55 &0.57 &0.56 &0.59 &0.52 \\
& & BO    &0.54 &0.44 &0.53 &0.45 &0.59 &0.50 \\
\bottomrule
\end{tabular}}
\caption{
Parameter estimates of Richards equation using GPPDE, GPC-I, BO-GPC and BO over 100 replicates with different number of observations and levels of noise when $L_m=3.2$ and $\beta=1.5$. The error variance is set to $b$ times the sample variance of the simulated state variable. Observations at 6 depths across 90 days are used. Estimates under fewer observations are also included.}
\end{table} 

\clearpage
\begin{figure}
\centering
\includegraphics[width=14cm]{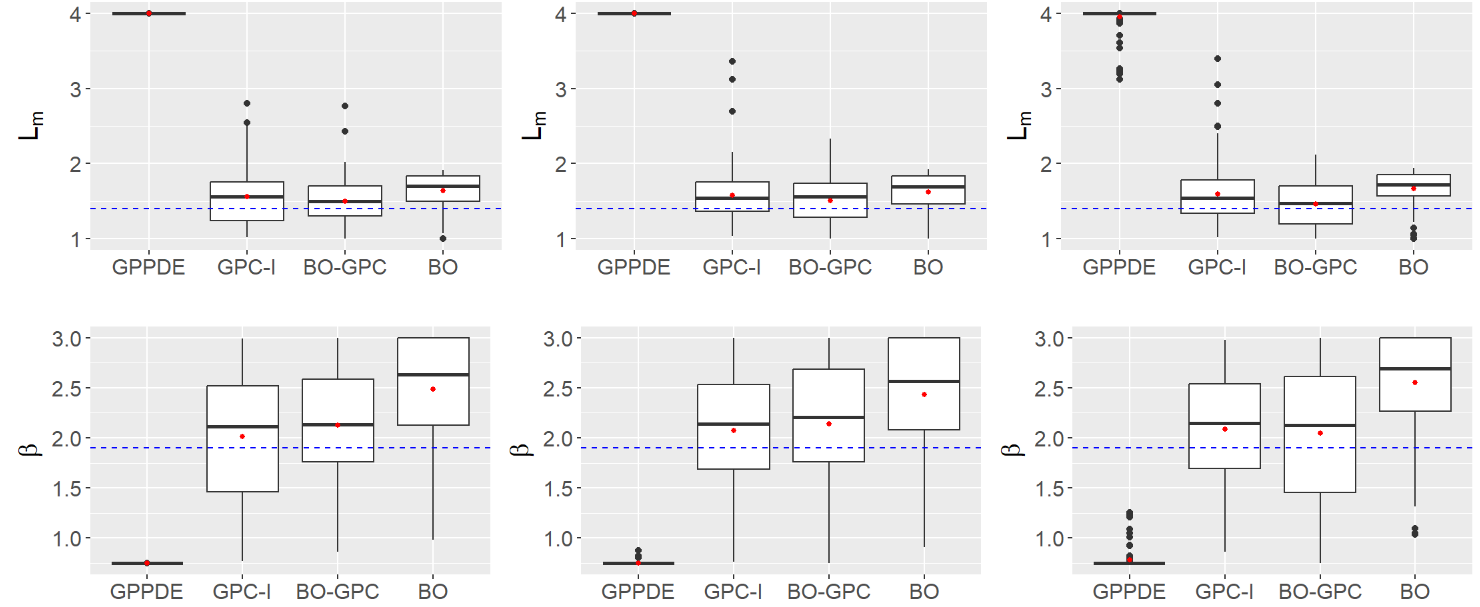}
\end{figure}

\begin{figure}
\centering
\includegraphics[width=14cm]{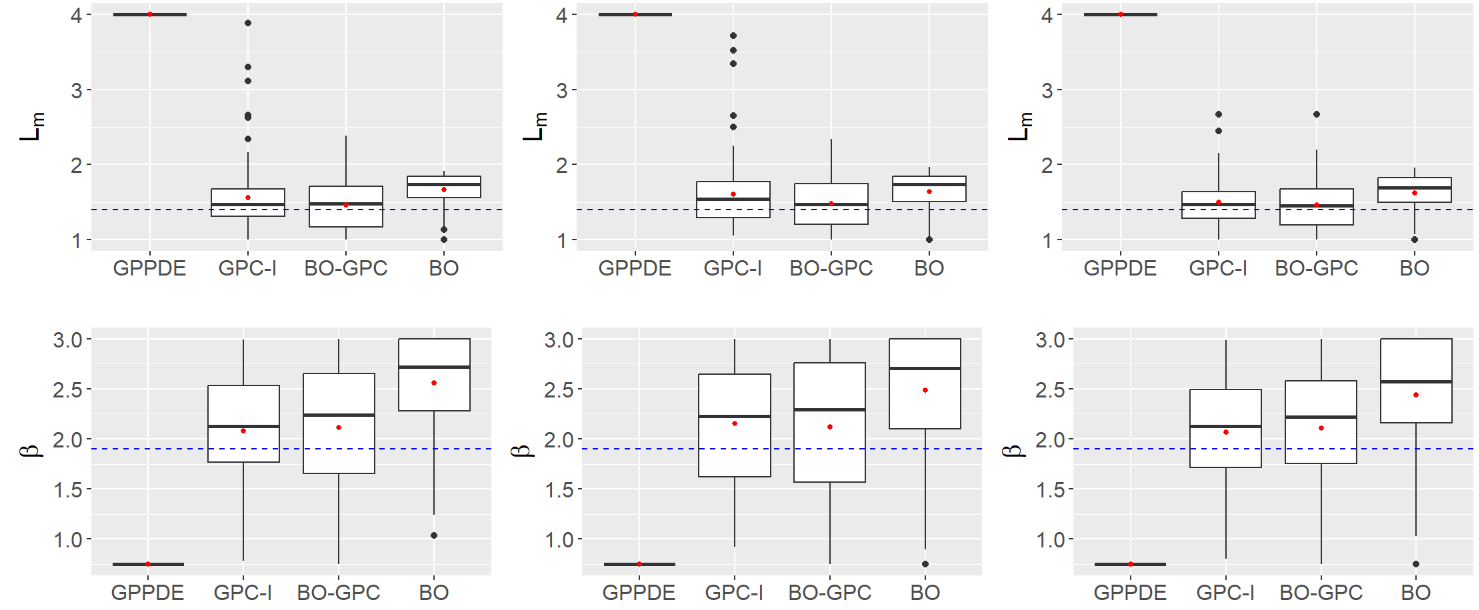}
\end{figure}

\begin{figure}
\centering
\includegraphics[width=14cm]{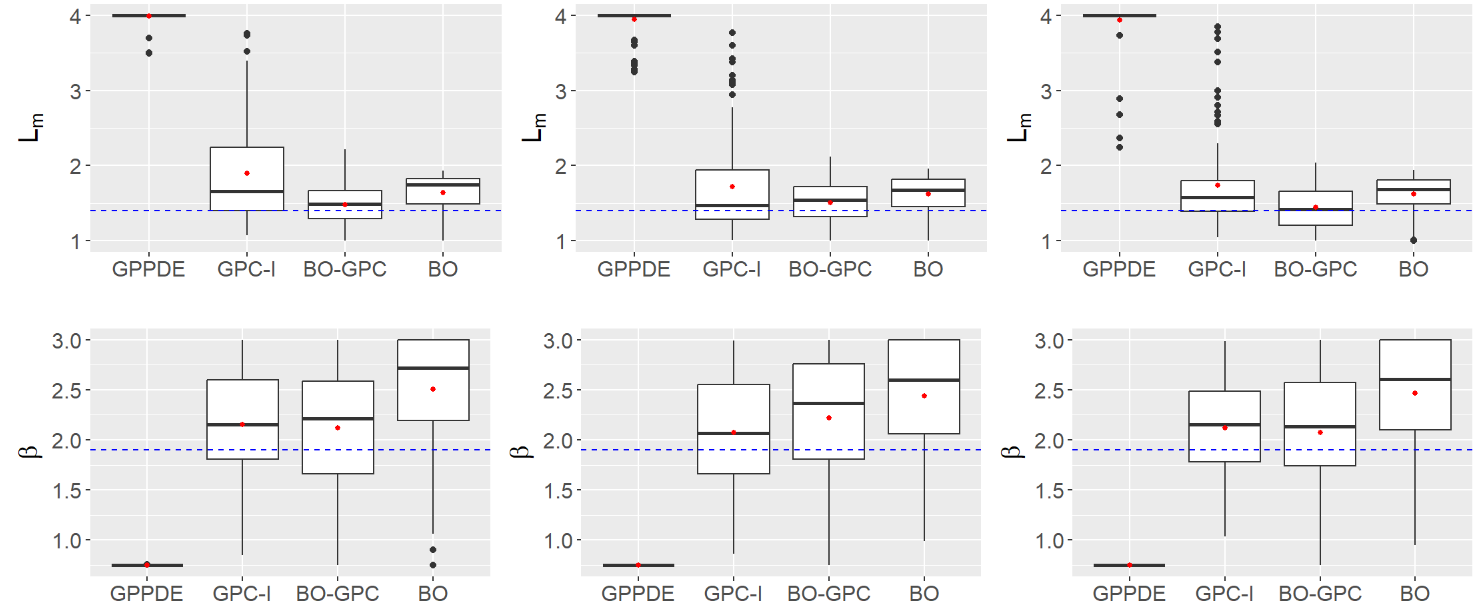}
\caption{Boxplots of the estimates obtained from GPPDE, GPC-I, BO-GPC and BO over 100 replicates at different error levels. Columns from left to right correspond to $b=$ 0.02, 0.05 and 0.1. Row 1 and 2 represents the estimates from 540 observations. Row 3 and 4 represents the estimates from 270 observations. Row 5 and 6 represents the estimates from 135 observations. The red dots represent the mean estimates from each method. The true parameter values are indicated by the blue dashed line ($L_m=1.4$, $\beta=1.9$). 
}
\end{figure}

\begin{figure}
\centering
\includegraphics[width=14cm]{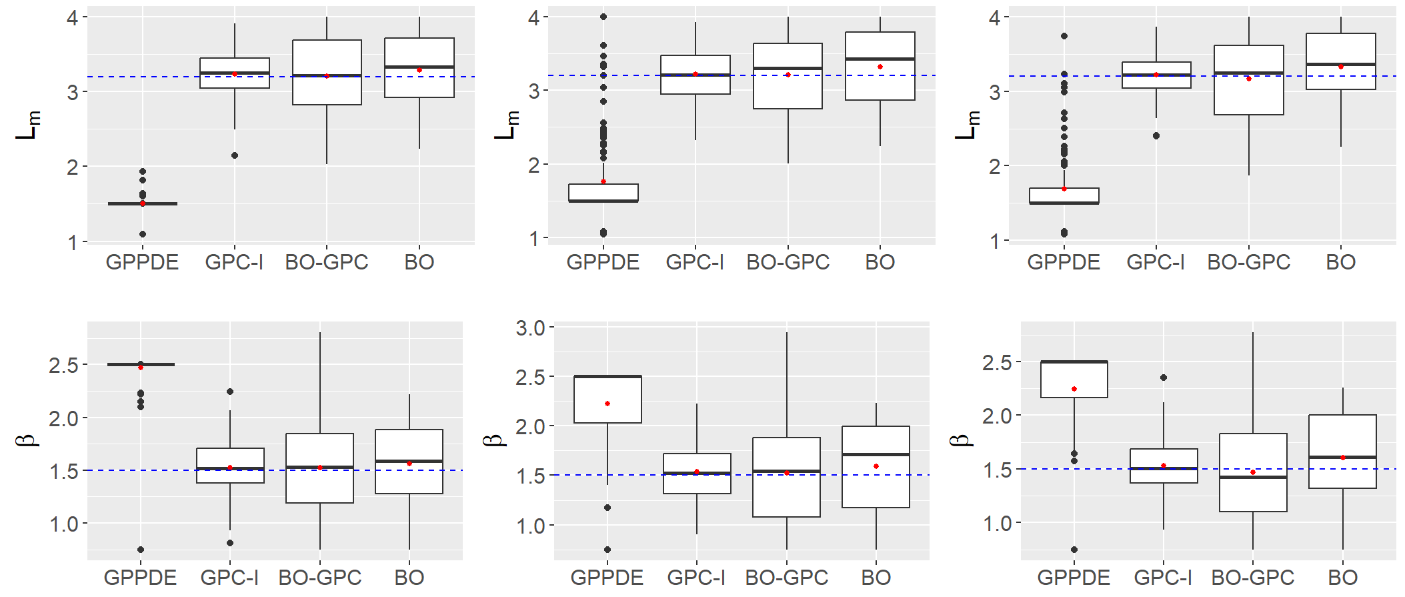}
\end{figure}

\begin{figure}
\centering
\includegraphics[width=14cm]{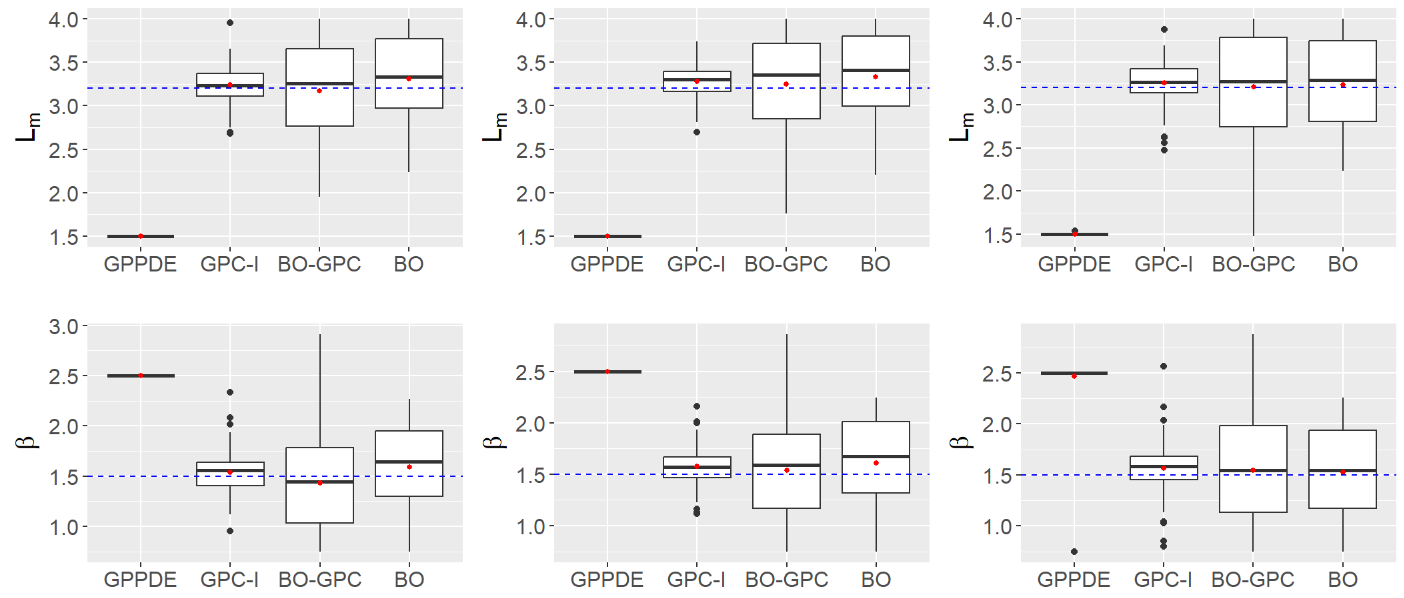}
\end{figure}

\begin{figure}
\centering
\includegraphics[width=14cm]{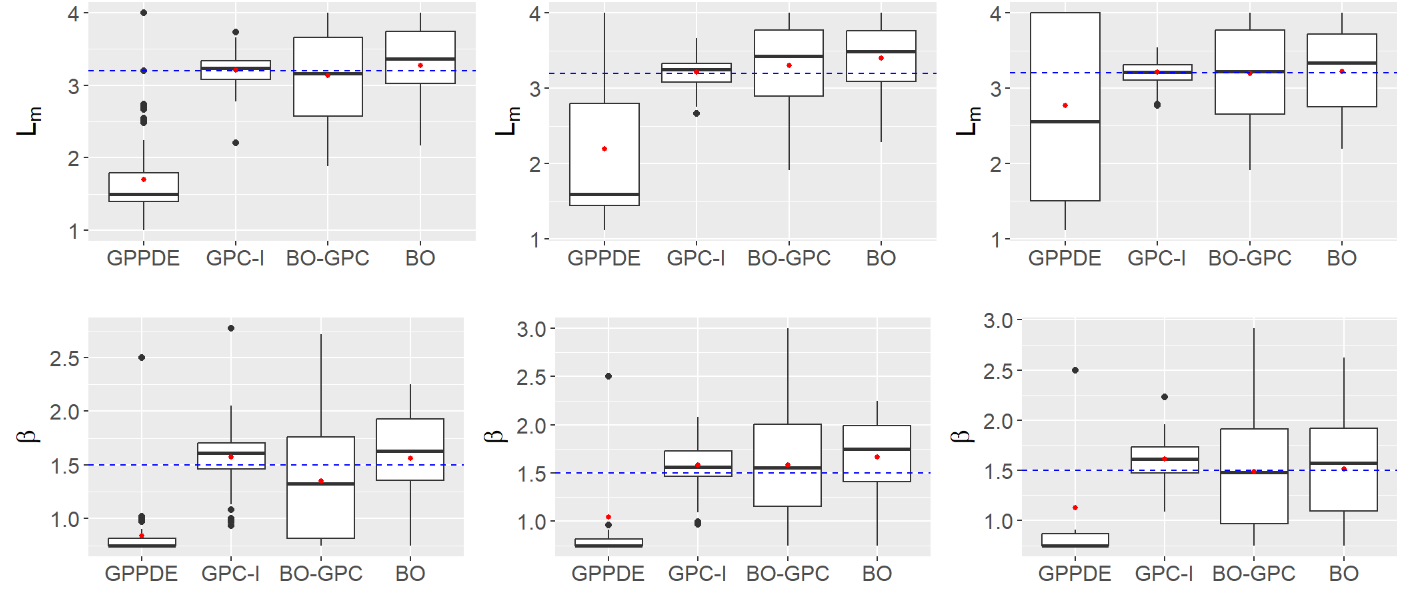}
\caption{Boxplots of the estimates obtained from GPPDE, GPC-I, BO-GPC and BO over 100 replicates at different error levels.  Columns from left to right correspond to $b=$ 0.02, 0.05 and 0.1. Row 1 and 2 represents the estimates from 540 observations. Row 3 and 4 represents the estimates from 270 observations. Row 5 and 6 represents the estimates from 135 observations. The red dots represent the mean estimates from each method. The true parameter values are indicated by the blue dashed line ($L_m=3.2$, $\beta=1.5$). 
}
\end{figure}

\end{document}

%% file: preamble.tex
\usepackage{float}

\DeclareMathOperator*\argmin{arg\,min}
\DeclareMathOperator*\argmax{arg\,max}

\newcommand{\btheta}{\bm{\theta}}

\newcommand{\blambda}{\bm{\lambda}}

\newcommand{\bA}{\bm{A}}

\newcommand{\bC}{\bm{C}}

\newcommand{\bK}{\bm{K}}

\newcommand{\cN}{\mathcal{N}}

\newcommand{\bX}{\bm{X}}

\newcommand{\bx}{\bm{x}}

\newcommand{\by}{\bm{y}}